\documentclass{bioinfo}

\usepackage{subfigure}

\usepackage{url}
\usepackage{multirow}
\usepackage{cuted}
\usepackage{arydshln}
\copyrightyear{2017} \pubyear{2017}

\access{Advance Access Publication Date: Day Month Year}
\appnotes{Manuscript Category}

\begin{document}
\firstpage{1}

\subtitle{Bioimage informatics}

\title[3D WaveUNet]{Neuron segmentation using 3D wavelet integrated encoder-decoder network}
\author[Qiufu Li \textit{et~al}.]{Qiufu Li$^{\text{\sfb 1,2,3}}$, Linlin Shen\,$^{\text{\sfb 1,2,3,4}*}$}
\address{$^{\text{\sf 1}}$Computer Vision Institute, College of Computer Science and Software Engineering, Shen zhen University, Shenzhen 518060, China, \\
$^{\text{\sf 2}}$AI Research Center for Medical Image Analysis and Diagnosis, Shenzhen University, Shenzhen 518060, China,\\
$^{\text{\sf 3}}$Guangdong Key Laboratory of Intelligent Information Processing, Shenzhen University, Shenzhen 518060, China,\\
$^{\text{\sf 4}}$Marshall Laboratory of Biomedical Engineering, Shenzhen University, Shenzhen 518060, China.}

\corresp{$^\ast$To whom correspondence should be addressed.}

\history{Received on XXXXX; revised on XXXXX; accepted on XXXXX}

\editor{Associate Editor: XXXXXXX}

\abstract{\textbf{Motivation:} 3D neuron segmentation is a key step for the neuron digital reconstruction,
   which is essential for exploring brain circuits and understanding brain functions.
   However, the fine line-shaped nerve fibers of neuron could spread in a large region, which brings great computational cost to the neuron segmentation.
   Meanwhile, the strong noises and disconnected nerve fibers bring great challenges to the task.\\
\textbf{Results:} In this paper, we propose a 3D wavelet and deep learning based 3D neuron segmentation method.
   The neuronal image is first partitioned into neuronal cubes to simplify the segmentation task.
   Then, we design 3D WaveUNet, the first 3D wavelet integrated encoder-decoder network, to segment the nerve fibers in the cubes;
   the wavelets could assist the deep networks in suppressing data noises and connecting the broken fibers.
   We also produce a \emph{Neu}ronal \emph{Cu}be \emph{Da}taset (NeuCuDa) using the biggest available annotated neuronal image dataset, BigNeuron, to train 3D WaveUNet.
   Finally, the nerve fibers segmented in cubes are assembled to generate the complete neuron, which is digitally reconstructed using an available automatic tracing algorithm.
   The experimental results show that our neuron segmentation method could completely extract the target neuron in noisy neuronal images.
   The integrated 3D wavelets can efficiently improve the performance of 3D neuron segmentation and reconstruction.\\
\textbf{Availability:} {\color{black}The data and codes for this work are available at https://github.com/LiQiufu/3D-WaveUNet.}\\
\textbf{Contact:} \href{qiufu_li_1988@163.com}{qiufu\_li\_1988@163.com}; \href{llshen@szu.edu.cn}{llshen@szu.edu.cn}\\
\textbf{Supplementary information:} Supplementary data are available at \textit{Bioinformatics}
online.}

\maketitle

\section{Introduction}
\begin{figure}[tbp]
\centering
\hspace{-5pt}
\subfigure{
		\label{fig_re_co_a}
		\includegraphics*[scale=0.45, viewport=29 596 147 822]{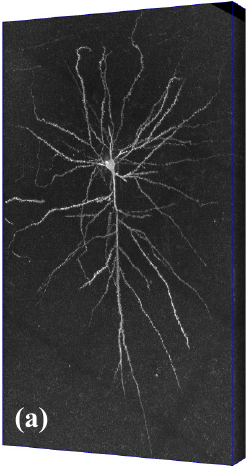}
	}\hspace{-4pt}
\subfigure{
		\label{fig_re_co_b}
		\includegraphics*[scale=0.45, viewport=169 596 289 822]{figures/reconstruction_comparison.pdf}
	}\hspace{-4pt}
\subfigure{
		\label{fig_re_co_c}
		\includegraphics*[scale=0.45, viewport=311 596 431 822]{figures/reconstruction_comparison.pdf}
	}\hspace{-4pt}
\subfigure{
		\label{fig_re_co_d}
		\includegraphics*[scale=0.45, viewport=453 596 567 822]{figures/reconstruction_comparison.pdf}
	}
   \caption{(a)~A noisy neuronal image with size of $291\times3298\times1881$ (z-y-x), containing a complete neuron.
   (b)~The manual reconstruction of the neuron.
   (c) An automatic reconstruction which incorrectly takes background noises as nerve fibers.
   (d) Another automatic reconstruction which overlooks the weak nerve fibers.}
\label{fig_reconstruction_comparison}
\end{figure}

Neuron reconstruction aims to establish a digital model of the neuron morphology structure by tracing nerve fibers in neuronal image,
which is of great significance to explore the brain microstructure and understand brain functions.
3D microscopic optical imaging technology, such as Micro-Optical Sectioning Tomography (MOST) \citep{li2010micro},
has established a foundation for the brain neuron reconstruction.
Tens of automatic or semi-automatic tracing algorithms \citep{xiao2013app2,liu2018automated,quan2016neurogps} have been developed to reconstruct the neurons in optical images.
However, the low signal-to-noise ratio and disconnected nerve fibers are usually big challenges to these algorithms, as Fig. \ref{fig_reconstruction_comparison} shows.

A natural way to improve the performance of existing neuron tracing algorithms is to extract the complete neuron from the noisy image first and then automatically reconstruct it.
Deep learning, which achieves excellent results in many fields, has also been applied into 3D neuron segmentation \citep{li2017deep,li20193d,jiang20203d,klinghoffer2020self,huang2020weakly}.
As nerve fibers can spread in a very large brain region,
a large neuronal image has to be processed for segmenting the complete neuron,
which leads to great computational cost for the deep networks based 3D neuron segmentation.
In addition, the recent researches \citep{qiufu_2020_CVPR,zhang2019making,zou2020delving} show that the sampling operations used in the deep networks result in aliasing effects,
which lead to noise propagation and break the basic object structures.
Due to the amount of noises available in the neuronal image and the neuron's fine line-shaped structure, 3D neuron segmentation is more sensitive to these aliasing effects.
Therefore, the current 3D deep networks based segmentation methods might not work well for the 3D neurons with fine line-shaped structure in the noisy neuronal images.

\subsection{Related works}
\label{sec_related_works}
\subsubsection{Neuron reconstruction}
\label{subsec_neuron_reconstruction}

The recent brain imaging technologies,
including Micro-Optical Sectioning Tomography (MOST) \citep{li2010micro}, fluorescence MOST (fMOST) \citep{gong2013continuously},
CLARITY \citep{chung2013clarity,chung2013structural}, Magnified Analysis of the Proteome (MAP) \citep{ku2016multiplexed},
and Stabilization under Harsh conditions via Intramolecular Epoxide Linkages to prevent Degradation (SHIELD) \citep{park2019protection}, etc.,
establish the foundation for neuron reconstruction of animal brain.

Various 3D data visualization software, such as Vaa3D \citep{peng2010v3d}, NeuroBlocks \citep{ai2015neuroblocks}, ManSegTool \citep{magliaro2017manual}, TeraVR \citep{wang2019teravr}, etc.,
have been developed to show the complex arborization fibers in neuronal images and assist the experienced technicians in tracing neurons.
BigNeuron project \citep{peng2015bigneuron}, initiated by Peng et al.,
releases hundreds of optical neuronal images and their high precision reconstructions traced by experienced experts.
More than $1\times10^5$ digitally reconstructed neurons are released on NeuroMorpho.Org \citep{ascoli2007neuromorpho},
while the corresponding neuronal images are not available.

It is impossible to manually reconstruct the millions of neurons in the whole brain image,
so various automatic neuron tracing algorithms have been designed.
The BigNeuron project \citep{peng2015bigneuron} collects the tens of these algorithms,
such as SnakeTracing \citep{wang2011broadly}, All-Path-Pruning (APP) \citep{peng2011automatic}, APP2 \citep{xiao2013app2},
NeuroGPSTree \citep{quan2016neurogps}, Rivulet \citep{liu2016rivulet}, Rivulet2 \citep{liu2018automated}, etc.,
and integrates them into the Vaa3D software \citep{peng2010v3d}.
Generally, these algorithms mathematically model neuron morphology and trace nerve fibers based on graph theory.
However, they would either classify background noise as nerve fibers or miss lots of nerve fibers, as Fig. \ref{fig_reconstruction_comparison} shows.
In other words, while the tracing algorithms are sensitive to image quality,
the existing neuronal imaging technologies are usually associated with low signal-to-noise ratio.

\subsubsection{Neuron segmentation}
\label{subsec_neuron_segmentation}
Segmenting the neuron before reconstruction is an effective approach to improve the performance of neuron tracing algorithms.
Inspired by the superior performance in computer vision, pattern recognition, and natural language processing, etc.,
deep learning has also been introduced into the 3D neuron segmentation \citep{li2017deep,li20193d,wang2019multiscale,jiang20203d,klinghoffer2020self,huang2020weakly}.
The first 3D residual deep network for neuron segmentation is proposed in \citep{li2017deep}.
The authors improve their method by applying multi-scale kernels and 3D U-Net \citep{wang2019multiscale}.
Because they take as input the original neuronal image, these deep networks require high computational cost.
\cite{li20193d} design 3D U-Net Plus to separate the target neuron from its surrounding nerve fibers.
As 3D U-Net Plus is trained and evaluated on the neuronal images with size of $32\times128\times128$,
the small nerve fibers may disappear during the image down-sampling.
The above deep network based neuron segmentation methods are not suitable for neurons with long-distance fibers.
Using ray-shooting model, \cite{jiang20203d} design dual channel Bidirectional Long Short-Term Memory (BLSTM) for 3D neuronal image segmentation.
The authors extract nerve fibers from the 2D neuronal image slice-by-slice,
which does not utilize the relevance between the nerve data in the neighbouring 2D neuronal slices.
To utilize the unlabeled 3D neuronal data, self-supervised and weakly supervised 3D neurons segmentation methods are studied \citep{klinghoffer2020self,huang2020weakly}.
In all of these works, the BigNeuron \citep{peng2015bigneuron} neuronal images are taken to test their methods.

\subsubsection{The aliasing effects in deep networks}
The recent studies \citep{qiufu_2020_CVPR,zhang2019making,zou2020delving} reveal the aliasing effects in the deep learning,
which are introduced by the commonly used sampling operations in the deep networks.
The aliasing effects result in noise accumulation and break the basic object structure in deep networks,
which are very harmful to the 3D neuron segmentation in noisy images.

To suppress the aliasing among the low-frequency and high-frequency components of the data in deep networks,
Richard \citep{zhang2019making} designs Anti-aliased CNNs by applying low-pass filters before the down-sampling in common Convolutional Neural Networks (CNNs).
Zou et al \citep{zou2020delving} propose an adaptive content-aware low-pass filtering layer to adapt the varying frequencies in different locations and feature channels.
These works only consider the low-frequency component of input data, which cannot extract and utilize the data details represented by the high-frequency components.
In \cite{qiufu_2020_CVPR}, the authors integrate discrete wavelet transforms (DWT) into deep networks
to separate the components of data in different frequency intervals.
They illustrate the usefulness of wavelet transforms in assisting deep networks in extracting robust features, keeping basic object structure, and recovering data details.

The above works suppress the aliasing effects in 2D deep networks, while they are not available to 3D deep networks for the 3D neuron segmentation.
The 3D neuronal images contain a lot of noises and the neurons are line-shaped,
which are more easily broken by the aliasing effecting in 3D deep networks.
Inspired by \cite{qiufu_2020_CVPR}, we try to introduce 3D wavelet transforms into deep learning to suppress the aliasing effects in 3D deep networks.
{\color{black}In the previous works, such as \cite{shi20173d,yang2019multi}, et al, 3D wavelet have been developed as pre-processing or post-processing tools for deep networks.}
Shi and Pun \citep{shi20173d} apply 3D wavelet to decompose hyperspectral image into various components wavelet domain,
and apply 3D CNNs on the components to extract features for the hyperspectral image classification.
Yang et al \citep{yang2019multi}, using 3D CNNs, predict the coefficients of hyperspectral image in 3D wavelet domain,
to reconstruct hyperspectral image with super-resolution.
{\color{black}In these works, only Haar wavelet was applied to evaluate their methods.
Because they do not integrate 3D wavelets into 3D deep networks,
they cannot suppress the aliasing effects in 3D deep networks.
In contrary, in this paper, we rewrite 3D wavelet transform as network layers in PyTorch,
which are applicable to various discrete wavelets (such as Haar, Cohen, and Daubechies wavelets),
and could be flexibly integrated into 3D deep networks to suppress the aliasing effects in networks.}

In this paper, we will rewrite 3D DWT/IDWT as general network layers, and design 3D wavelet integrated encoder-decoder networks,
to improve the 3D neuron segmentation and reconstruction in noisy images.

\subsection{Contributions}
In this paper, to completely segment the neuron and efficiently improve its reconstruction,
we present a 3D wavelet and deep learning based 3D neuron segmentation method.
Firstly, to simplify the task, we partition the neuronal image into small cubes.
Then, we design a 3D wavelet integrated encoder-decoder network (3D WaveUNet) to segment the nerve fibers in the cubes.
As far as we know, 3D WaveUNet is the first 3D wavelet integrated deep network.
In 3D WaveUNet, 3D Discrete Wavelet Transform (3D DWT) and 3D Inverse DWT (3D IDWT) are applied to down-sample and up-sample 3D neuronal data.
While 3D DWT help the encoders in maintaining the fine structure of neurons and suppressing the noise propagation,
3D IDWT could recover the neuron details in the decoders.
In addition, based on the biggest available annotated neuronal image dataset, BigNeuron \citep{peng2015bigneuron},
we produce a \emph{Neu}ronal \emph{Cu}be \emph{Da}taset (NeuCuDa) to train and evaluate the 3D WaveUNet.
Finally, we assemble the nerve fibers segmented in cubes according to the cube locations to get the complete segmented neuron,
and reconstruct it using the existing automatic tracing algorithms, such as All-Path Pruning 2.0 (APP2) \citep{xiao2013app2}.
In summary:
\vspace{-0pt}
\begin{enumerate}
\renewcommand{\labelenumi}{\arabic{enumi}.}
\setlength{\topsep}{-0ex}
\setlength{\itemsep}{-0.0ex}
  \item We propose a 3D wavelet and deep learning based 3D neuron segmentation method, to segment neurons in large size 3D neuronal images.
  \item We rewrite 3D DWT/IDWT as general network layers, and design 3D WaveUNet by applying 3D DWT and IDWT as the sampling operations in 3D encoder-decoder network,
    which is the first 3D wavelet integrated deep network.
  \item We produce a neuronal cube dataset, NeuCuDa, to train and evaluate the 3D WaveUNet.
  Experimental results show that 3D WaveUNet efficiently improve the performance of 3D neuron segmentation and reconstruction in the noisy images.
\end{enumerate}

\section{Method}
\subsection{The method pipeline}
\label{sec_method_pipeline}
\begin{figure}[tbp]
\centering
\includegraphics*[scale=0.4, viewport=7 568 586 752]{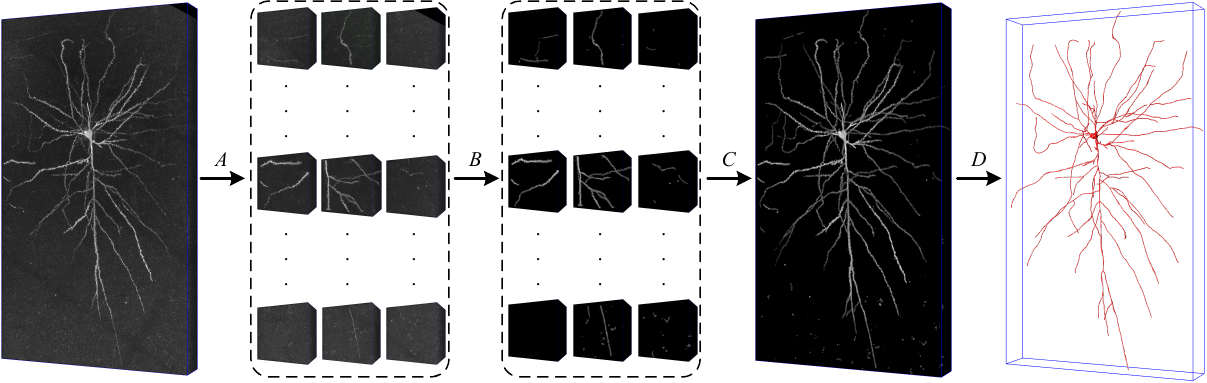}
   \caption{The pipeline of our 3D neuron segmentation method.
   \emph{A}: Partition.
   \emph{B}: Segmentation.
   \emph{C}: Assembling.
   \emph{D}: Reconstruction.}
\label{fig_method_pipeline}
\end{figure}
The pipeline of our 3D neuron segmentation method is illustrated in Fig. \ref{fig_method_pipeline} and summarized as below:
\begin{enumerate}
\setlength{\topsep}{-0ex}
\setlength{\itemsep}{-0.0ex}
  \renewcommand{\labelenumi}{\arabic{enumi})}
  \item \emph{Partition}.
  The neurons with long-distance nerve fibers could spread in a large brain region,
  which result in high computational cost for the 3D neuron segmentation in the images with large size.
  Therefore, the neuronal image is partitioned into small cubes to simplify the segmentation task.
  We set the z-y-x-size (depth-height-width) of the cubes as $32\times128\times128$
  \footnote {The typical voxel resolution is $1\mu \text{m} \times 0.35\mu \text{m} \times 0.35 \mu \text{m}$ for brain imaging technologies, such as MOST.},
  which is a compromise between the connectivity of nerve fibers in the neuronal cubes and the computational complexity of the following segmentation.

  \item \emph{Segmentation}.
  Using the well-trained 3D WaveUNet, we segment the nerve fibers in the cubes.
  The 3D WaveUNet is an encoder-decoder network integrated with 3D wavelet, which are trained on the neuronal cube dataset, NeuCuDa.
  The network architecture and NeuCuDa are described in Sec. \ref{sec_3D_waveunet} and Sec. \ref{sec_NeuCuDa}, respectively.

  \item \emph{Assembling}.
  According to the cube locations in the neuronal image,
  we assemble the segmented nerve fibers to complete the 3D neuron segmentation.

  \item \emph{Reconstruction}.
  Based on segmented neuronal image, we reconstruct the neuron using APP2 \citep{xiao2013app2} algorithm.
  Considering its low computational complexity, we choose APP2 as the benchmark algorithm in this paper.

\end{enumerate}

\subsection{3D WaveUNet}
\label{sec_3D_waveunet}
The aliasing effects could significantly affect the performance of deep network based 3D neuron segmentation,
due to the fine line-shaped structure of nerve fibers and interference of noises.
To suppress the aliasing effects in the 3D deep networks,
we design 3D wavelet integrated encoder-decoder network (3D WaveUNet),
the first 3D wavelet integrated deep network, for 3D neuron segmentation.

{\color{black}
In current, encoder-decoder networks, such as U-Net and 3D U-Net, are widely used in medical image segmentation \citep{ronneberger2015u,li20193d}.
Although ResNets \citep{he2016deep} and Transformer \citep{dosovitskiy2020image,touvron2020training} are hot-spots in the image vision,
they are hardly applied in medical image processing.
Because ResNets and Transformer are pretrained using a large dataset (such as ImageNet) before they are taken as backbones in the deep networks for image segmentation,
and, for 3D medical image processing, it is very hard to collect such large dataset.
The proposed 3D DWT and IDWT layers are natural substitutes for the down-sampling and up-sampling operations in encoder and decoder of 3D U-Net.
Therefore, we take 3D U-Net as the basic architecture, to evaluate the proposed 3D DWT/IDWT layers in 3D neuron segmentation.
}

\subsubsection{3D DWT/IDWT layers}
Wavelets \citep{daubechies1992ten} are powerful analysis tools used in the fields of data denoising \citep{donoho1995noising,donoho1994ideal},
image compression \citep{kim1997embedded,wu2009hyperspectral,bahce2020compression}, etc.,
and 2D wavelets have been introduced into deep learning
\citep{huang2017wavelet,duan2017sar,liu2018multi,williams2018wavelet,guan2019wavelet,yoo2019photorealistic,liu2020wavelet,qiufu_2020_CVPR,yang2019multi,shi20173d}.
{\color{black}Alike 2D wavelets in deep learning, 3D wavelets are applied as pre-processing or post-processing tools in deep networks.
\cite{shi20173d} apply 3D wavelet to decompose hyperspectral image into various components wavelet domain,
and extract features from the components using 3D CNNs to classify the hyperspectral image.
\cite{yang2019multi} predict the coefficients of hyperspectral image in 3D wavelet domain using 3D CNNs,
to reconstruct hyperspectral image with super-resolution.
These works do not apply 3D wavelets into the design of 3D deep networks,
cannot exploit the efficiency of wavelets to process the internal 3D feature maps in deep networks.}

We rewrite 3D Discrete Wavelet Transform (DWT) and Inverse DWT (IDWT) as general network layers, to integrate 3D wavelets into deep networks.
For a given 3D neuronal data $\textbf{\emph{X}}\in\mathbb{R}^{d\times m \times n}$,
and eight filters of a 3D discrete wavelet, i.e., one low-pass filter $\text{f}_{lll}$ and seven high-pass filters
$\text{f}_{llh}, \text{f}_{lhl}, \text{f}_{lhh}, \text{f}_{hll}, \text{f}_{hlh}, \text{f}_{hhl}, \text{f}_{hhh}$,
3D DWT decomposes the data into one low-frequency component $\textbf{\emph{X}}_{lll}$ and seven high-frequency components
$\textbf{\emph{X}}_{llh}, \textbf{\emph{X}}_{lhl}, \textbf{\emph{X}}_{lhh},
\textbf{\emph{X}}_{hll}, \textbf{\emph{X}}_{hlh}, \textbf{\emph{X}}_{hhl}, \textbf{\emph{X}}_{hhh}$,
where
\begin{align}
\label{eq_DWT}
\textbf{\emph{X}}_{c_0c_1c_2} = (\downarrow2)(\text{f}_{c_0c_1c_2}\ast\textbf{\emph{X}}),\quad c_0, c_1, c_2  \in \{l, h\}
\end{align}
and $\ast, (\downarrow2)$ denote the 3D convolution and naive down-sampling, respectively.
In theory, the size of every component is $1/2$ size of $\textbf{\emph{X}}$ in every dimension,
i.e.,
\begin{align}
\textbf{\emph{X}}_{c_0c_1c_2} \in \mathbb{R}^{\lfloor\frac{d}{2}\rfloor\times\lfloor\frac{m}{2}\rfloor\times\lfloor\frac{n}{2}\rfloor},
\quad c_0, c_1, c_2  \in \{l, h\}.
\end{align}
Therefore, $d, m, n$ are usually even numbers.

\begin{figure}[tbp]
	\centering
	\includegraphics*[scale=0.725, viewport=91 107 405 356]{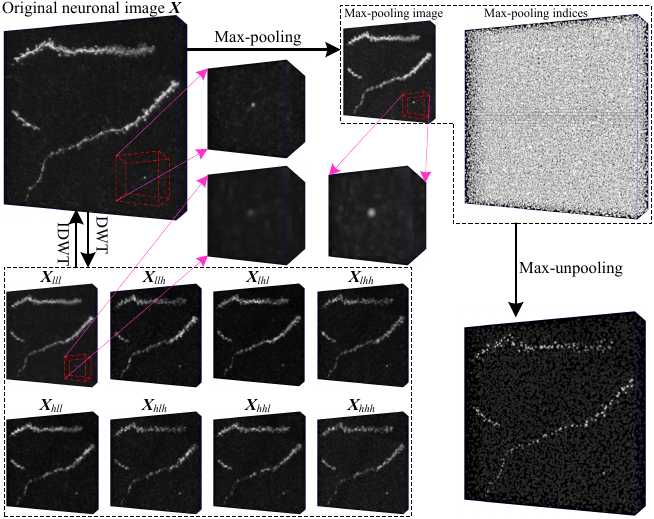}
	\caption{The comparison of 3D max-pooling/max-unpooling and 3D DWT/IDWT.
            Max-pooling would magnify the noise in the neuronal data,
            while noise in the low-frequency component $\emph{\textbf{X}}_{lll}$ is suppressed by 3D DWT, as the three enlarged regions show.
            When max-unpooling recovers the cube size, it normally breaks the nerve fibers.
            In contrast, IDWT could completely recover the neuronal data using the components of DWT decomposition.}
            \label{fig_down_sampling}
\end{figure}

Using the dual filters $\tilde{\text{f}}_{c_0c_1c_2},~c_0, c_1, c_2  \in \{l, h\}$, 3D IDWT reconstructs the original data $\textbf{\emph{X}}$ based on the eight components,
\begin{align}
\label{eq_IDWT}
\textbf{\emph{X}} &= \sum_{c_0,c_1,c_2\in\{l,h\}}\tilde{\text{f}}_{c_0c_1c_2}\ast(\uparrow2)\textbf{\emph{X}}_{c_0c_1c_2},
\end{align}
where $(\uparrow2)$ denotes the naive up-sampling operation.
The dual filters of orthogonal 3D wavelet are the same with the original filters,
\begin{align}
\label{eq_filter}
\tilde{\text{f}}_{c_0c_1c_2} = \text{f}_{c_0c_1c_2},~c_0, c_1, c_2  \in \{l, h\}.
\end{align}

Eqs. (\ref{eq_DWT}) and (\ref{eq_IDWT}) present the forward propagations of 3D DWT and IDWT,
where the naive down-sampling and up-sampling operations are explained in Supplementary (Sec. \ref{sec_naive_sampling}).
It is onerous to deduce the gradients for the backward propagations from these expressions.
Fortunately, the modern deep learning framework PyTorch \citep{paszke2017automatic}
could automatically deduce the gradients for tensor arithmetics.
We have rewritten 3D DWT and IDWT as general network layers in PyTorch,
which will be publicly available for other researchers.
One can flexibly design end-to-end 3D wavelet integrated deep networks using these layers.

For the noisy neuronal cube $\textbf{\emph{X}}$, the random noise mostly show up in its high-frequency components,
while the basic nerve fiber structure is presented by the low-frequency one.
Therefore, 3D DWT and IDWT could be used to denoise the neuronal cube while maintaining the structure of nerve fiber at the same time.
DWT halves the size of the neuronal data, and IDWT recovers it,
which are good substitutes for the commonly used sampling operations in the 3D deep networks.
They could be used to reduce the aliasing effects in the 3D deep networks,
and improve the performance of 3D neuron segmentation and reconstruction.
Fig. \ref{fig_down_sampling} compares the results of 3D DWT/IDWT and max-pooling/max-unpooling on a neuronal cube.

\subsubsection{The architecture of 3D WaveUNets}
\begin{figure}[bpt]
	\centering
	\includegraphics*[scale=0.4, viewport=25 187 569 439]{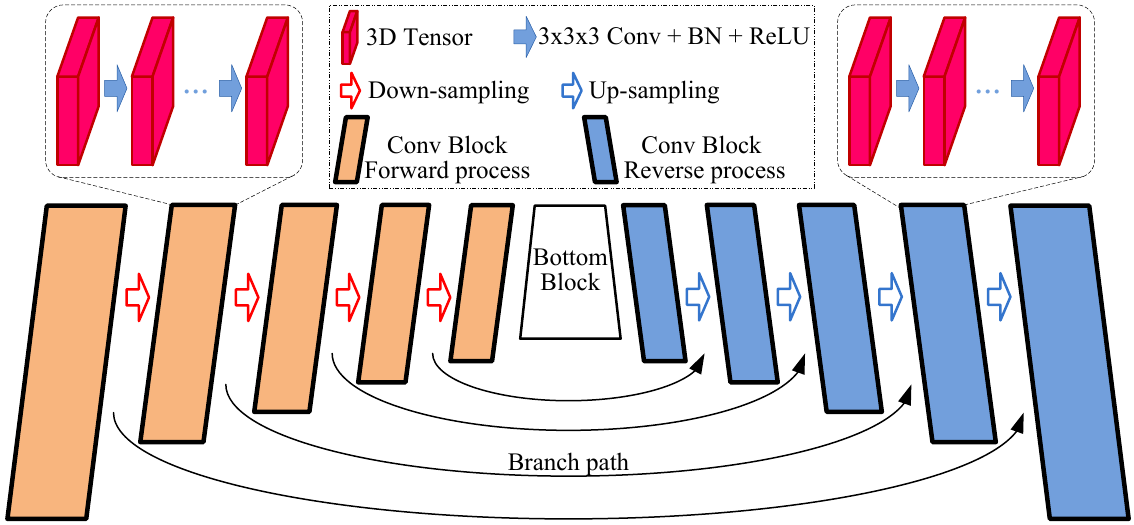}
	\caption{The encoder-decoder architecture.}\label{fig_encoder_decoder}
\end{figure}
In this section, we design 3D WaveUNet using 3D DWT and IDWT
to increase neuron segmentation performance for better reconstruction.

As Fig. \ref{fig_encoder_decoder} shows,
the encoder-decoder architecture could be decomposed into several nested \textbf{dual structure}s with a \textbf{bottom block}.
The dual structure consists of \textbf{forward process} and \textbf{reverse process} connected via a \textbf{branch path}.
While the forward process contains a serial of 3D convolutions followed by a down-sampling operation,
the reverse process is an up-sampling operation followed by a number of 3D convolutions.
The reverse process up-samples the feature maps,
and exploits the data transmitted from the forward process via the branch path.
The bottom block is variant in different encoder-decoder architectures.
3D U-Net \citep{ronneberger2015u} takes a convolution block as the bottom block,
while 3D U-Net Plus \citep{li20193d} employs a 3D ASPP \citep{chen2018encoder_deeplabv3+}.
We denote the all connected forward processes and reverse processes,
together with the bottom block, as \textbf{main path},
and denote the data flowing through it as \textbf{mainstream}.

\begin{figure}[tbp]
	\centering
	\subfigure[DS-PU, 3D version of dual structure used in SegNet \citep{badrinarayanan2017segnet}]
	{\includegraphics*[scale=0.5, viewport=89 679 506 808]{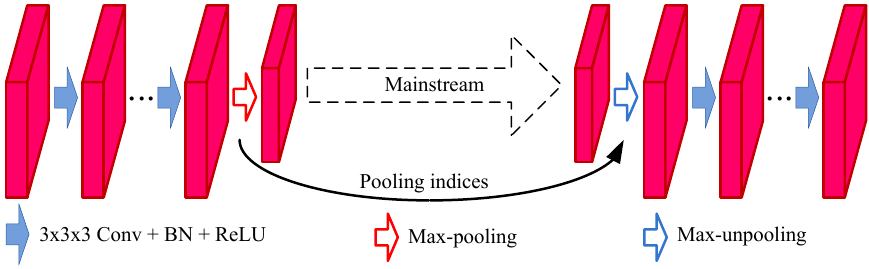}
	\label{fig_dual_structure_a}}\\
	\subfigure[DS-PDc used in 3D U-Net Plus \citep{li20193d} and 3D U-Net \citep{cciccek20163d}]
	{\includegraphics*[scale=0.5, viewport=89 522 506 651]{figures/dual_structures.pdf}
	\label{fig_dual_structure_b}}\\
	\subfigure[DS-ScIn, 3D version of dual structure used in DeepLabV3+ \citep{chen2018encoder_deeplabv3+}]
	{\includegraphics*[scale=0.5, viewport=75 367 519 495]{figures/dual_structures.pdf}
	\label{fig_dual_structure_c}}
    \caption{The 3D versions of the commonly used dual structures.}
	\label{fig_dual_structure}
\end{figure}

\begin{figure}[tbp]
	\centering
	\subfigure[WADS-DDc (DWT, Deconvolute)]
	{\includegraphics*[scale=0.5, viewport=89 598 506 727]{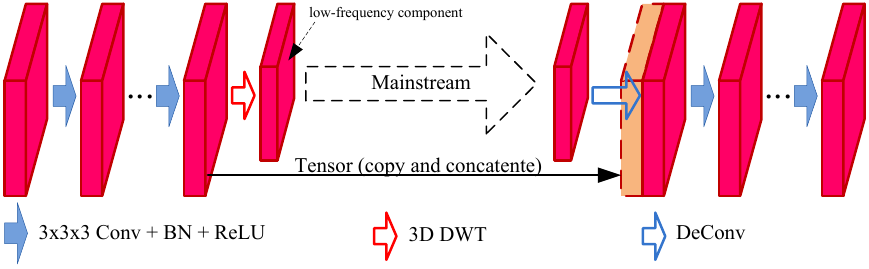}
	\label{fig_dual_structure_wavelet_a}}\\
	\subfigure[WADS-DIn (DWT, Interpolate)]
	{\includegraphics*[scale=0.5, viewport=89 441 506 569]{figures/dual_structures_wavelet.pdf}
	\label{fig_dual_structure_wavelet_b}}\\
	\subfigure[WADS-DI (DWT, IDWT)]
	{\includegraphics*[scale=0.5, viewport=89 269 506 398]{figures/dual_structures_wavelet.pdf}
	\label{fig_dual_structure_wavelet_c}}\\
	\subfigure[WADS-DIDn (DWT, IDWT; Denoise)]
	{\includegraphics*[scale=0.5, viewport=89 113 506 243]{figures/dual_structures_wavelet.pdf}
	\label{fig_dual_structure_wavelet_d}}\\
    \caption{The 3D wavelet integrated dual structures (WADS).}
	\label{fig_dual_structure_wavelet}
\end{figure}
Fig. \ref{fig_dual_structure} shows the 3D versions of dual structures used in previous encoder-decoder networks
\citep{ronneberger2015u,badrinarayanan2017segnet,chen2018encoder_deeplabv3+,cciccek20163d,li20193d}.
DS-PU, Dual Structure with max-Pooling and max-Unpooling, is a 3D version of dual structure used in SegNet \citep{badrinarayanan2017segnet}.
As Fig. \ref{fig_dual_structure_a} shows, DS-PU applies max-pooling in its forward process to reduce the feature map resolution
and transmits the pooling indices to the reverse process for feature map up-sampling via max-unpooling.
However, the max-pooling operation would accumulate the data noise,
and the max-unpooling cannot restore the lost details, as Fig. \ref{fig_down_sampling} shows.
Fig. \ref{fig_dual_structure_b} shows DS-PDc, Dual Structure with max-Pooling and Deconvolution, used
in 3D U-Net \citep{cciccek20163d} and 3D U-Net Plus \citep{li20193d}.
While DS-PDc applies the max-pooling for the down-sampling in its forward process,
it uses deconvolution for the up-sampling in the reverse process.
DS-PDc copies the feature maps from its forward process to reverse process via the branch path,
then concatenates it with the up-sampled feature map in reverse process
to recover the low-level information in mainstream.
DS-ScIn, shown in Fig. \ref{fig_dual_structure_c}, has the same structure with DS-PDc,
while it applies Strided-convolution and linear Interpolation for down-sampling and up-sampling, respectively.
In DS-PDc and DS-ScIn, the data tensor injected from their forward processes to reverse processes
contains redundant information including noises, which could interfere with 3D neuron segmentation.

We design four 3D WAvelet based Dual Structure (WADS) by replacing the sampling operations with 3D DWT/IDWT,
as Fig. \ref{fig_dual_structure_wavelet} shows.
In comparison,
WADS-DDc (Fig. \ref{fig_dual_structure_wavelet_a}) and WADS-DIn (Fig. \ref{fig_dual_structure_wavelet_b}) are designed
by replacing the down-sampling operations in DS-PDc and DS-ScIn with 3D DWT, respectively.
In their forward processes, while the feature map is decomposed by 3D DWT into eight components,
only the low-frequency one is kept to extract the robust high-level features, and the seven high-frequency components are abandoned.
WADS-DI (Fig. \ref{fig_dual_structure_wavelet_c}) and WADS-DIDn (Fig. \ref{fig_dual_structure_wavelet_d})
apply 3D DWT and 3D IDWT for down-sampling and up-sampling in their forward and reverse processes, respectively.
In the forward process, while feature map is decomposed by 3D DWT,
the low-frequency component is kept to extract the robust high-level features,
and the seven high-frequency components are transmitted, via the branch path, to reverse process for feature map up-sampling in 3D IDWT.
WADS-DIDn utilizes a denoising block to filter the high-frequency components;
the denoising block is implemented using hard shrinkage, which is shown in the Supplementary.

Using the seven dual structures, we design seven 3D encoder-decoder networks, including four wavelet integrated ones (3D WaveUNets), for neuron segmentation.
They are named as 3D U-Net($x$), $x \in \{\text{PU, PDc, ScIn}\}$ and 3D WaveUNet($y$), $y \in \{\text{DDc, DIn, DI, DIDn}\}$, respectively.
The detail configurations for these deep networks are shown in the Supplementary (Sec. \ref{sec_configuration_3D_WaveUNet}).
\begin{figure}[!t]
	\centering
	\includegraphics*[scale=0.88, viewport=11 382 277 830]{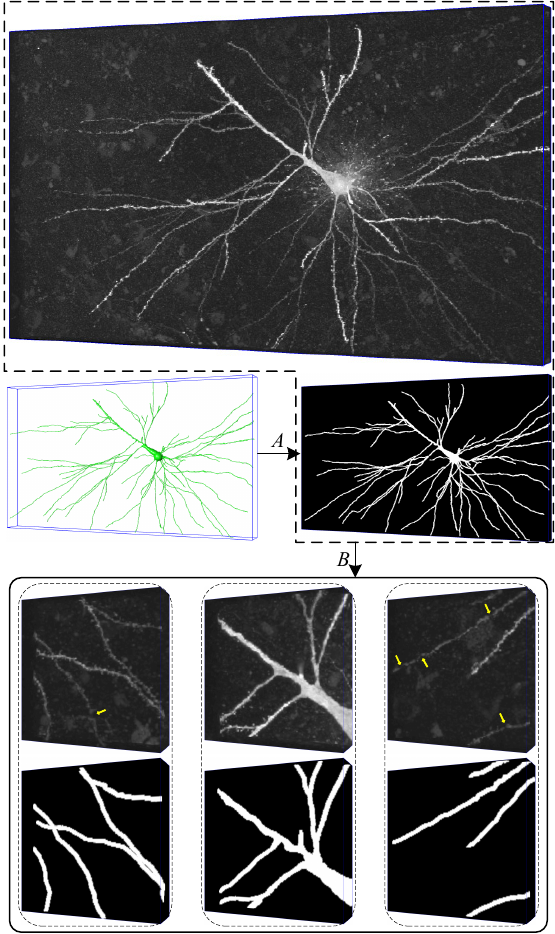}
	\caption{A neuronal image and three cubes cut from it. \emph{A}: labeling, \emph{B}: cutting.
            The cubes are corrupted by random noises;
            some weak nerve fibers in the cubes are disconnected, as marked by the yellow arrows.
            The 3D label matrices are shown below the cubes.
            }
    \label{fig_neuronal_cubes}
\end{figure}
\subsection{The neuronal cube dataset}
\label{sec_NeuCuDa}
As we partition the large neuron images into small cubes for neuron segmentation,
it is crucial to collect an appropriate dataset to train and evaluate 3D WaveUNets.
We here introduce the \emph{Neu}ronal \emph{Cu}be \emph{Da}taset, NeuCuDa, generated from the BigNeuron images.
{\color{black}The neuronal images in BigNeuron are captured from various specials, including silk moth, frog, mouse, etc,
using different imaging equipments.
The diversity of neuronal data helps to train robust deep networks.}
The procedure to produce NeuCuDa is shown in Fig. \ref{fig_neuronal_cubes} and summarized as below:
\vspace{-0pt}
\begin{enumerate}
\renewcommand{\labelenumi}{\arabic{enumi})}
\setlength{\topsep}{-0ex}
\setlength{\itemsep}{-0.0ex}
  \item \emph{Selection}.
        {\color{black}
        In BigNeuron, some neuronal reconstructions do not provide full information (such as nerve fiber radius) of the neurons.
        For such neuronal images, we cannot correctly label the fibers for segmentation.
        Therefore, we exclude these neuronal images in designing NeuCuDa.}
        From the remaining 89 neuronal images,
        we randomly choose 61 images to generate the neuronal cubes for the training of 3D deep networks,
        and 28 images to generate the testing cubes.
        The 28 testing images will also be used to evaluate the performance of reconstruction method proposed in Sec. \ref{sec_method_pipeline}.
  \item \emph{Labeling}. For each of the chosen 89 (61 + 28) neuronal images, we generate a 3D 0-1 label matrix according to its standard digital reconstruction.
        Every voxel in the images is assigned to 0 (background) or 1 (nerve fiber).
        An example neuronal image with its 3D label matrix is shown in Fig. \ref{fig_neuronal_cubes}.
  \item \emph{Cutting}. Neuronal cubes with size of $32\times128\times128$ are randomly cut from the 3D image.
        Meanwhile, their label cubes are cut from the 3D label matrix.
        Fig. \ref{fig_neuronal_cubes} shows three example neuronal cubes.
\end{enumerate}
\vspace{-0pt}
NeuCuDa contains 19251 and 7132 neuronal cubes for the training and testing of 3D WaveUNet, respectively.

\section{Results}
\label{sec_experiments}
\begin{table*}
\scriptsize
    \caption{Segmentation and reconstruction results.}\label{Tab_segmentation_reconstruction}
    \begin{center}
    \setlength{\tabcolsep}{0.5mm}{
    \begin{tabular}{c|c|ccc|cccc|c||ccc||ccc||c|c}\cline{1-18}
    \multirow{2}{*}{Architecture} &Dual& \multicolumn{3}{c|}{Encoder} & \multicolumn{4}{c|}{Decoder} &\multirow{2}{*}{wavelet}
    & \multicolumn{3}{c||}{Segmentation (IoU)} & \multicolumn{3}{c||}{Reconstruction} & \multirow{2}{*}{Parameters}& \multirow{2}{*}{\color{black}GFLOPs}\\\cline{3-9}\cline{11-16}
                            & Structure  & \underline{P}ool & \underline{S}tr-\underline{c}onv &  \underline{D}WT
                            & \underline{U}npool & \underline{D}e\underline{c}onv & \underline{In}terpolate &  \underline{I}DWT && bg$^a$ &ner\_fib&mean&ESA&DSA&PDS&  \\\cline{1-18}\cline{1-18}
    baseline (APP2)         &--&--&--&--&--&--&--&--&--&--&--&--&9.3942&14.3641&0.3573&--&--\\\cline{1-18}
    \multirow{3}{*}{3D U-Net}&DS-PU& $\surd $&  &  & $\surd $&  &  &    &none &99.04 &45.55 &72.30 &2.5444 &    7.3450 &    0.2200 & \textbf{0.17}~$\times10^6$&\textbf{1.3771}~$\times10^9$\\\cline{2-18}
                        &DS-PDc& $\surd $&  &  &   &$\surd $&  &    &none &99.29 &53.65 &76.43 &2.4046 &    7.1799 &    0.1950 & 0.20~$\times10^6$ &{1.8010}~$\times10^9$\\\cline{2-18}
                        &DS-ScIn& &$\surd $  &  & &  & $\surd $ &    &none &99.21 &52.96 &76.09 &2.3614 &    6.9439 &    0.1968 & 0.20~$\times10^6$ &{1.7990}~$\times10^9$\\\cline{1-18}
    \multirow{12}{*}{3D WaveUNet}&\multirow{3}{*}{WADS-DDc}& & &\multirow{3}{*}{$\surd $} & &\multirow{3}{*}{$\surd$}& &    &haar &\textbf{99.32} &\textbf{54.80} &\textbf{77.06} &2.3272 &    6.8841 &    0.1962  &\multirow{3}{*}{0.20~$\times10^6$}&\multirow{3}{*}{{1.8022}~$\times10^9$}\\
                                                & & & &                          & & &                         &    &ch2.2 &99.32 &54.58 &76.95 &2.2495 &    6.5712 &    0.1956  &\\
                                                & & & &                          & & &                         &    &db3 &90.27 &54.76 &77.02 &2.3391  &   7.0144  &   0.1931  &\\\cline{2-18}
                             &\multirow{3}{*}{WADS-DIn}& & &\multirow{3}{*}{$\surd $} & & &\multirow{3}{*}{$\surd $} &    &haar &99.20 &52.59 &75.90 &2.5247 &    7.1783 &    0.2083 &\multirow{3}{*}{0.20~$\times10^6$}&\multirow{3}{*}{{1.8022}~$\times10^9$}\\
                                                & & & &                          & &                          & &    &ch2.2 &99.27 &52.89 &76.08 &2.0288 &    6.2569 &    0.1922 &\\
                                                & & & &                          & &                          & &    &db2 & 99.20 &52.67 &75.94 &2.2197  &   6.6692  &   0.1947  & \\\cline{2-18}
                             &\multirow{3}{*}{WADS-DI}& & &\multirow{3}{*}{$\surd $} & & & &\multirow{3}{*}{$\surd$}    &haar &99.30 &54.31 &76.81 &2.0588  &   6.4043  &   0.1866  &\multirow{3}{*}{\textbf{0.17}~$\times10^6$}&\multirow{3}{*}{{1.8532}~$\times10^9$}\\
                                                & & & &                          & & & &                            &ch4.4 &99.28 &54.46 &76.87 &2.1726 &    6.3609 &    0.1938  &\\
                                                & & & &                          & & & &                            &db4 &99.22 &53.06 & 76.14&2.0676  &   6.2566  &   \textbf{0.1857} & \\\cline{2-18}
                             &\multirow{3}{*}{WADS-DIDn$^b$}& & &\multirow{3}{*}{$\surd $} & & & &\multirow{3}{*}{$\surd$}    &haar   &99.30 &54.76 &77.03 &2.0336 &    6.0730 &    0.1868  &\multirow{3}{*}{\textbf{0.17}~$\times10^6$}&\multirow{3}{*}{{1.8556}~$\times10^9$}\\
                                                & & & &                          & & & &                            &ch4.4 &99.26 &54.78 &77.02 &2.2067  &   6.6038 &    0.1914  &\\
                                                & & & &                          & & & &                            &db4 & 99.24& 53.22& 76.23&\textbf{1.9973}   &  \textbf{6.0173} &    0.1897&  \\\cline{1-18}\cline{1-18}
    \end{tabular}}
    \end{center}
    \vspace{-0pt}
    \hspace{.35cm}$^a$ ``bg'' and ``ner\_fib'' are short for ``background'' and ``nerve fiber'', respectively. \par
    \hspace{.35cm}$^b$ ``Dn'' stands for ``\underline{D}e\underline{n}oising block''.
\end{table*}

We first train and evaluate the seven 3D deep networks on NeuCuDa.
Then, using the well-trained networks, we segment and reconstruct the neurons in the 28 test neuronal images with the proposed method in Sec. \ref{sec_method_pipeline}.

\subsection{3D neuronal cube segmentation}
On the 19251 training cubes, we train the 3D U-Nets and 3D WaveUNets for 30 epochs, using stochastic gradient descent (SGD).
The training is driven by cross-entropy loss with weights 1, 5 for ``background'' and ``nerve fiber'' voxels, to address the imbalances in the cubes.
We initially take a learning rate of 0.1, and reduce it following a polynomial decay for each iteration.
The batch size, momentum, and weight decay are set as 32, 0.9, and 0.0001, respectively.
The segmentation performances on the 7132 test cubes are shown in Table \ref{Tab_segmentation_reconstruction}.
We train 3D WaveUNets when various wavelets are used,
and for each network architecture, the results of three wavelets are shown in Table \ref{Tab_segmentation_reconstruction}.
In the table, ``haar'' stands for the Haar wavelet,
while ``db$p$'' and ``ch$p.\tilde{p}$'' stand for Daubechies with approximation order $p$ and Cohen wavelet with orders $p,\tilde{p}$, respectively.
The filters of these wavelets are shown in Supplementary (Sec. \ref{sec_3D_wavelet_filters}).

3D U-Net(PU), applying max-pooling/max-unpooling,
performs the worst on neuronal cube segmentation among the seven deep networks.
This result suggests that the pair of sampling operations is not suitable for neuron segmentation, which matches our analysis.
Comparing the segmentation results of 3D U-Net(PDc) and 3D WaveUNet(DDc),
one can find that DWT performs better than max-pooling,
when deconvolution is taken as up-sampling operation;
3D WaveUNet(DDc) achieves the best segmentation performance (mIoU, $77.06\%$).

Among the seven networks, the segmentation performances of 3D U-Net(ScIn) and 3D WaveUNet(DIn) are only better than that of 3D U-Net(PU),
which indicates that the linear interpolation used in their decoder is not a good up-sampling operation for 3D neuron segmentation.

3D WaveUNet(DI) and 3D WaveUNet(DIDn), which apply 3D DWT and IDWT for neuron data down-sampling and up-sampling,
perform also well for neuronal cube segmentation.
Comparing their results, we conclude that denoising on high-frequency components could slightly improve the performance of neuronal cube segmentation.
The segmentation performance of 3D WaveUNet(DIDn) is close to that of 3D WaveUNet(DDc), although the convolutions in decoder of 3D WaveUNet(DDc) employs more parameters.
In summary, the 3D wavelets improve the performance of neuron segmentation in 3D noisy images.

{\color{black}
In Table \ref{Tab_segmentation_reconstruction},
the last column lists the No. of multiply-add operations required by various 3D deep networks, for processing a neuronal cube with size of $32\times128\times128$.
3D U-Net(PU) applies the simple down-sampling and up-sampling operations, i.e., Max-pooling and Max-unpooling, which takes the least No. of multiply-add operations (1.3771 G).
3D U-Net(PDc) and 3D U-Net(ScIn) apply strided-convolution and DeConv for their down-sampling and up-sampling,
which require 1.8010 G and 1.7990 G multiply-add operations, respectively.
In 3D WaveUNets, the No. of operations required by wavelet transforms are less than 3\% of that of 3D U-Net(PDc) and 3D U-Net(ScIn),
while the segmentation and reconstruction performances of 3D WaveUNets are superior.}

{\color{black}
For segmentation of the 28 test neuronal images, Table \ref{Tab_segmentation_time} shows the average processing time of various 3D deep networks, on an Nvidia V100 GPU.
In average, our method needs about two seconds to segment a 3D neuronal image.
}
\begin{table}
\color{black}
\scriptsize
    \caption{Average processing times of various 3D deep networks for segmentation of the 28 test neuronal images.}\label{Tab_segmentation_time}
    \begin{center}
    \setlength{\tabcolsep}{1.25mm}{
    \begin{tabular}{c|c|c|c|c|c|c|c}\hline
    \multirow{1}{*}{Architecture}&\multicolumn{3}{c|}{3D U-Net}&\multicolumn{4}{c}{3D WaveUNet}\\\hline
    Dual Structure &PU&	PDc&	ScIn&	DDc&	DIn&	DI&	DIDn\\\hline\hline
    Times (seconds)	&\textbf{1.4810}&	1.4657&	1.4703&	1.7276&	1.7751&	2.0980&	2.1678\\\hline
    \end{tabular}}
    \end{center}
\end{table}
\begin{figure*}[bpt]
\centering
	\subfigure
	{\includegraphics*[scale=0.9, viewport= 14 635 114 830]{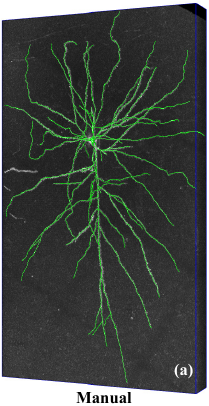}
	\label{fig_reconstruction_neuron20_a}}
	\subfigure
	{\includegraphics*[scale=0.9, viewport=127 635 227 830]{figures/reconstruction_neuron20.pdf}
	\label{fig_reconstruction_neuron20_b}}
	\subfigure
	{\includegraphics*[scale=0.9, viewport=241 635 341 830]{figures/reconstruction_neuron20.pdf}
	\label{fig_reconstruction_neuron20_c}}
	\subfigure
	{\includegraphics*[scale=0.9, viewport=354 635 454 830]{figures/reconstruction_neuron20.pdf}
	\label{fig_reconstruction_neuron20_d}}
	\subfigure
	{\includegraphics*[scale=0.9, viewport=468 635 568 830]{figures/reconstruction_neuron20.pdf}
	\label{fig_reconstruction_neuron20_e}}\\\vspace{-10pt}
	\subfigure
	{\includegraphics*[scale=0.9, viewport= 14 437 114 632]{figures/reconstruction_neuron20.pdf}
	\label{fig_reconstruction_neuron20_f}}
	\subfigure
	{\includegraphics*[scale=0.9, viewport=127 437 227 632]{figures/reconstruction_neuron20.pdf}
	\label{fig_reconstruction_neuron20_g}}
	\subfigure
	{\includegraphics*[scale=0.9, viewport=241 437 341 632]{figures/reconstruction_neuron20.pdf}
	\label{fig_reconstruction_neuron20_h}}
	\subfigure
	{\includegraphics*[scale=0.9, viewport=354 437 454 632]{figures/reconstruction_neuron20.pdf}
	\label{fig_reconstruction_neuron20_i}}
	\subfigure
	{\includegraphics*[scale=0.958, viewport=468 437 568 632]{figures/reconstruction_neuron20.pdf}
	\label{fig_reconstruction_neuron20_j}}\\
    \caption{The reconstructions of various methods. (a) The reconstruction manually traced by experts.
    (b) - (c) Two reconstructions of APP2 on the original neuronal image, with different parameters.
    (d) - (f) The reconstructions of APP2 on the images segmented by the three 3D U-Nets.
    (g) - (j) The reconstructions of APP2 on the images segmented by the four 3D WaveUNets.
    {\color{black}In the reconstructing experiments, we tune two parameters of APP2, i.e., background threshold and length threshold.
    For the neuron reconstructions shown in Fig. \ref{fig_reconstruction_neuron20_b}, we set the two parameters as 10 and 5, respectively, which are the default values in APP2,
    For the one shown in Fig. \ref{fig_reconstruction_neuron20_c}, we set both two parameters as 1.
    For all the neuron reconstructions traced on the images segmented by 3D deep networks, we set both of the APP2 parameters as 1.}
    }
	\label{fig_reconstruction_neuron20}
\end{figure*}

\subsection{3D neuron reconstruction}
After segmenting the nerve fibers in the partitioned cubes, we assemble 3D neurons in the 28 test neuronal images, and reconstruct them using APP2, as described in Sec. \ref{sec_method_pipeline}.
We evaluate the accuracy of automatic reconstructions by comparing them with the manual one,
using three metrics (lower is better), i.e., ``entire structure average (ESA)'', ``different structure average (DSA)'', and ``percentage of different structures (PDS)'',
proposed in  \citep{peng2010automatic} by Peng et al.
Table \ref{Tab_segmentation_reconstruction} shows the three mean metric values of the 28 neurons reconstructed using APP2 on the segmentation results of 3D U-Nets and 3D WaveUNets.

The three mean metric values (9.3942, 14.3641, and 0.3573) of APP2 on the 28 original noisy images are taken as baseline.
From Table \ref{Tab_segmentation_reconstruction}, one can find that the reconstruction performance on the segmented images of the seven deep networks is clearly superior to the baseline,
which indicates that our method significantly improve the 3D neuron reconstruction.
Generally, with the application of 3D DWT and IDWT in the encoder and decoder of the deep networks,
the reconstruction performance is getting better and better,
and 3D WaveUNet(DIDn) with wavelet ``db4'' achieves the best reconstruction performance (1.9973, 6.0173, and 0.1897).
In summary, our 3D wavelet and deep network based method efficiently improve the automatic reconstruction performance for 3D neurons in the noisy images.

{\color{black}
Comparing the segmentation and reconstruction results shown in Table \ref{Tab_segmentation_reconstruction},
we find that the deep network with better segmentation performance may not lead to the better neuron reconstruction,
mainly due to the label noises in the dataset NeuCuda.
According to the manual reconstructions of neurons in BigNeuron,
we design an automatic software to label the neuronal cubes of NeuCuDa.
However, during the implementation, some label noises are introduced into the dataset.
As Fig. \ref{fig_neuronal_cubes} shows, some background voxels surrounding the nerve fibers are incorrectly labeled as neuronal voxels.
These label noises result in the inconsistency between the neuron segmentation and its final reconstruction. }

In Supplementary (Sec. \ref{sec_results_various_tracing_algorithms}),
to better illustrate the effectiveness of 3D wavelet integrated deep networks on 3D neuron segmentation and reconstruction,
we apply various automatic tracing algorithms to reconstruct the test images segmented by the different 3D deep networks integrated with or without 3D wavelets.

Fig. \ref{fig_reconstruction_neuron20} 
visually show the reconstructions of an example noisy neuronal image.
In Fig. \ref{fig_reconstruction_neuron20}, the original size of the image is $291\times3298\times1881$ (z-y-x),
while the previous deep network based 3D neuron segmentation methods \citep{li2017deep,li20193d,wang2019multiscale} cannot process the neuronal images with such a large size.
Fig. \ref{fig_reconstruction_neuron20_a} shows the manual reconstruction of the neuron, which perfectly reflect the neuron morphology.
Fig. \ref{fig_reconstruction_neuron20_b} and Fig. \ref{fig_reconstruction_neuron20_c} show the reconstructions of APP2 with different parameters on the original neuronal image.
The following three subfigures show the reconstructions of APP2 on the neurons segmented by the three 3D U-Nets,
and the last four subfigures show that of the four 3D WaveUNets.
{\color{black}While the segmentation process takes the three 3D U-Nets about $4.2$ seconds,
it takes the four 3D WaveUNets about $4.9, 5.0, 6.1$, and $6.4$ seconds, respectively.
}

On the original noisy image, the automatic tracing algorithm APP2 either stop tracing weak nerve fibers (Fig. \ref{fig_reconstruction_neuron20_b})
or over-reconstruct strong noises as nerve fibers (Fig. \ref{fig_reconstruction_neuron20_c}).
From Fig. \ref{fig_reconstruction_neuron20}, one can find that the reconstructions of APP2 on the segmented image are superior to that on the original image,
which illustrates the efficiency of our 3D neuron reconstruction method.
The 3D U-Nets either classify ``background'' voxels as ``nerve fiber'' voxels or do not extract complete nerve fibers.
Therefore, their segmented neuronal images contain much background noise (Fig. \ref{fig_reconstruction_neuron20_d} and Fig. \ref{fig_reconstruction_neuron20_e}),
or the reconstructions are not complete (Fig. \ref{fig_reconstruction_neuron20_d} and Fig. \ref{fig_reconstruction_neuron20_f}), as marked by the yellow arrows.
The 3D WaveUNets obtain cleaner segmentation and more complete reconstruction,
which justifies the effectiveness of 3D wavelets in suppressing noise and keeping nerve fiber structure.
Comprehensively, the 3D WaveUNet(DIDn), which utilizes DWT, IDWT, and denoising block, achieves the best segmentation and reconstruction results.

{\color{black}We compare our method with three more recent deep networks \citep{li2017deep,li20193d,jiang20203d}.
Table \ref{Tab_reconstruction_performance_various_method} lists the results.
For ``3D Residual Net'', we compute the three mean metric values (3.2892, 5.8900, and 0.3367) of APP2 performance on 17 BigNeuron images published in \citep{li2017deep}.
For ``3D U-Net Plus'', we reconstruct the neurons in 28 test neuronal images using the well-trained 3D U-Net Plus and reconstruction method,
and the reconstruction performance is 33.5020, 40.3151, and 0.5539.
On the 28 test neuronal images, the three metrics of ``Bidirectional Long Short-Term Memory (BLSTM)'' are 6.7582, 13.0149, and 0.3646.
While the DSA value (5.8900) of 3D Residual Net is slightly better than that of our method,
the network employs significantly more parameters (11.41 M) than our 3D WaveUNets (0.17 M).
The performance of 3D U-Net Plus is even inferior to the baseline (shown in Table \ref{Tab_segmentation_reconstruction}).
As the method resizes the neuronal image into $32\times128\times128$ before segmentation,
it cannot completely segment the neurons in large size neuronal images.
BLSTM segment every 2D neuronal slice to denoise the 3D image, which performs badly for disconnected nerve fibers.
Comprehensively, among these four methods, our 3D WaveUNets perform best in terms of both metrics and network complexity.
}
\begin{table}
\centering
\color{black}
\scriptsize
\caption{Reconstruction performance of different 3D deep networks.}\label{Tab_reconstruction_performance_various_method}
    \setlength{\tabcolsep}{1.25mm}{
\begin{tabular}{c|ccc|c}\cline{1-5}
\multirow{2}{*}{Network} & \multicolumn{3}{c|}{Reconstruction}   & \multirow{2}{*}{Parameters} \\\cline{2-4}
                        &ESA&DSA&PDS& \\\hline\hline
3D WaveUNet (Ours) &\textbf{1.9973} & 6.0173 & \textbf{0.1897}& \textbf{0.17} $\times10^6$ \\\hline
3D Residual Net \citep{li2017deep}    &3.2892 & \textbf{5.8900} & 0.3367& 11.41 $\times10^6$ \\\hline
3D U-Net Plus \citep{li20193d}     &33.5020 & 40.3151 & 0.5539& 3.60 $\times10^6$ \\\hline
BLSTM \citep{jiang20203d} &6.7582&13.0149&0.3646&0.60$\times10^6$\\\hline
\end{tabular}}
\end{table}

%
%

\section{Conclusion}
\label{sec_conclusions}
We propose a 3D wavelet and deep learning based neuron segmentation method, which could improve the reconstruction performance for 3D neurons with large size.
We rewrite 3D DWT/IDWT as general network layers, and design the first 3D wavelet integrated encoder-decoder network, 3D WaveUNet, for 3D neuron segmentation.
The 3D wavelet transforms could suppress the noise propagation and keep nerve fiber structure during the deep network inference,
thus improve the performance of 3D neuron segmentation and reconstruction in noisy neuronal images.

In future, we will study the 3D neuron instance segmentation for images containing multiple neurons,
and extend the 3D wavelet integrated deep networks to more 3D medical image applications.


\section*{Funding}

This work was supported by the National Natural Science Foundation of China [62006156, 91959108],
the Science and Technology Project of Guangdong Province [2018A050501014],
and the Science Foundation of Shenzhen [JSGG20180508152022006].\vspace*{-12pt}

%
%

\bibliographystyle{model2-names}
\bibliography{3D_WaveUNet}

\begin{thebibliography}{54}
\expandafter\ifx\csname natexlab\endcsname\relax\def\natexlab#1{#1}\fi
\providecommand{\url}[1]{\texttt{#1}}
\providecommand{\href}[2]{#2}
\providecommand{\path}[1]{#1}
\providecommand{\DOIprefix}{doi:}
\providecommand{\ArXivprefix}{arXiv:}
\providecommand{\URLprefix}{URL: }
\providecommand{\Pubmedprefix}{pmid:}
\providecommand{\doi}[1]{\href{http://dx.doi.org/#1}{\path{#1}}}
\providecommand{\Pubmed}[1]{\href{pmid:#1}{\path{#1}}}
\providecommand{\bibinfo}[2]{#2}
\ifx\xfnm\relax \def\xfnm[#1]{\unskip,\space#1}\fi
\bibitem[{Ai-Awami et~al.(2015)Ai-Awami, Beyer, Haehn, Kasthuri, Lichtman,
  Pfister and Hadwiger}]{ai2015neuroblocks}
\bibinfo{author}{Ai-Awami, A.K.}, \bibinfo{author}{Beyer, J.},
  \bibinfo{author}{Haehn, D.}, \bibinfo{author}{Kasthuri, N.},
  \bibinfo{author}{Lichtman, J.W.}, \bibinfo{author}{Pfister, H.},
  \bibinfo{author}{Hadwiger, M.}, \bibinfo{year}{2015}.
\newblock \bibinfo{title}{Neuroblocks--visual tracking of segmentation and
  proofreading for large connectomics projects}.
\newblock \bibinfo{journal}{IEEE transactions on visualization and computer
  graphics} \bibinfo{volume}{22}, \bibinfo{pages}{738--746}.
\bibitem[{Ascoli et~al.(2007)Ascoli, Donohue and
  Halavi}]{ascoli2007neuromorpho}
\bibinfo{author}{Ascoli, G.A.}, \bibinfo{author}{Donohue, D.E.},
  \bibinfo{author}{Halavi, M.}, \bibinfo{year}{2007}.
\newblock \bibinfo{title}{Neuromorpho. org: a central resource for neuronal
  morphologies}.
\newblock \bibinfo{journal}{Journal of Neuroscience} \bibinfo{volume}{27},
  \bibinfo{pages}{9247--9251}.
\bibitem[{Badrinarayanan et~al.(2017)Badrinarayanan, Kendall and
  Cipolla}]{badrinarayanan2017segnet}
\bibinfo{author}{Badrinarayanan, V.}, \bibinfo{author}{Kendall, A.},
  \bibinfo{author}{Cipolla, R.}, \bibinfo{year}{2017}.
\newblock \bibinfo{title}{Segnet: A deep convolutional encoder-decoder
  architecture for image segmentation}.
\newblock \bibinfo{journal}{IEEE transactions on pattern analysis and machine
  intelligence} \bibinfo{volume}{39}, \bibinfo{pages}{2481--2495}.
\bibitem[{Bahce and Bayazit(2020)}]{bahce2020compression}
\bibinfo{author}{Bahce, C.G.}, \bibinfo{author}{Bayazit, U.},
  \bibinfo{year}{2020}.
\newblock \bibinfo{title}{Compression of geometry videos by 3d-speck wavelet
  coder}.
\newblock \bibinfo{journal}{VISUAL COMPUTER} .
\bibitem[{Chen et~al.(2018)Chen, Zhu, Papandreou, Schroff and
  Adam}]{chen2018encoder_deeplabv3+}
\bibinfo{author}{Chen, L.C.}, \bibinfo{author}{Zhu, Y.},
  \bibinfo{author}{Papandreou, G.}, \bibinfo{author}{Schroff, F.},
  \bibinfo{author}{Adam, H.}, \bibinfo{year}{2018}.
\newblock \bibinfo{title}{Encoder-decoder with atrous separable convolution for
  semantic image segmentation}, in: \bibinfo{booktitle}{Proceedings of the
  ECCV}, pp. \bibinfo{pages}{801--818}.
\bibitem[{Chung and Deisseroth(2013)}]{chung2013clarity}
\bibinfo{author}{Chung, K.}, \bibinfo{author}{Deisseroth, K.},
  \bibinfo{year}{2013}.
\newblock \bibinfo{title}{Clarity for mapping the nervous system}.
\newblock \bibinfo{journal}{Nature methods} \bibinfo{volume}{10},
  \bibinfo{pages}{508}.
\bibitem[{Chung et~al.(2013)Chung, Wallace, Kim, Kalyanasundaram, Andalman,
  Davidson, Mirzabekov, Zalocusky, Mattis, Denisin
  et~al.}]{chung2013structural}
\bibinfo{author}{Chung, K.}, \bibinfo{author}{Wallace, J.},
  \bibinfo{author}{Kim, S.Y.}, \bibinfo{author}{Kalyanasundaram, S.},
  \bibinfo{author}{Andalman, A.S.}, \bibinfo{author}{Davidson, T.J.},
  \bibinfo{author}{Mirzabekov, J.J.}, \bibinfo{author}{Zalocusky, K.A.},
  \bibinfo{author}{Mattis, J.}, \bibinfo{author}{Denisin, A.K.}, et~al.,
  \bibinfo{year}{2013}.
\newblock \bibinfo{title}{Structural and molecular interrogation of intact
  biological systems}.
\newblock \bibinfo{journal}{Nature} \bibinfo{volume}{497},
  \bibinfo{pages}{332--337}.
\bibitem[{{\c{C}}i{\c{c}}ek et~al.(2016){\c{C}}i{\c{c}}ek, Abdulkadir,
  Lienkamp, Brox and Ronneberger}]{cciccek20163d}
\bibinfo{author}{{\c{C}}i{\c{c}}ek, {\"O}.}, \bibinfo{author}{Abdulkadir, A.},
  \bibinfo{author}{Lienkamp, S.S.}, \bibinfo{author}{Brox, T.},
  \bibinfo{author}{Ronneberger, O.}, \bibinfo{year}{2016}.
\newblock \bibinfo{title}{3d u-net: learning dense volumetric segmentation from
  sparse annotation}, in: \bibinfo{booktitle}{International conference on
  medical image computing and computer-assisted intervention},
  \bibinfo{organization}{Springer}. pp. \bibinfo{pages}{424--432}.
\bibitem[{Daubechies(1992)}]{daubechies1992ten}
\bibinfo{author}{Daubechies, I.}, \bibinfo{year}{1992}.
\newblock \bibinfo{title}{Ten lectures on wavelets}.
  volume~\bibinfo{volume}{61}.
\newblock \bibinfo{publisher}{Siam}.
\bibitem[{Donoho(1995)}]{donoho1995noising}
\bibinfo{author}{Donoho, D.L.}, \bibinfo{year}{1995}.
\newblock \bibinfo{title}{De-noising by soft-thresholding}.
\newblock \bibinfo{journal}{IEEE transactions on information theory}
  \bibinfo{volume}{41}, \bibinfo{pages}{613--627}.
\bibitem[{Donoho and Johnstone(1994)}]{donoho1994ideal}
\bibinfo{author}{Donoho, D.L.}, \bibinfo{author}{Johnstone, J.M.},
  \bibinfo{year}{1994}.
\newblock \bibinfo{title}{Ideal spatial adaptation by wavelet shrinkage}.
\newblock \bibinfo{journal}{biometrika} \bibinfo{volume}{81},
  \bibinfo{pages}{425--455}.
\bibitem[{Dosovitskiy et~al.(2020)Dosovitskiy, Beyer, Kolesnikov, Weissenborn,
  Zhai, Unterthiner, Dehghani, Minderer, Heigold, Gelly
  et~al.}]{dosovitskiy2020image}
\bibinfo{author}{Dosovitskiy, A.}, \bibinfo{author}{Beyer, L.},
  \bibinfo{author}{Kolesnikov, A.}, \bibinfo{author}{Weissenborn, D.},
  \bibinfo{author}{Zhai, X.}, \bibinfo{author}{Unterthiner, T.},
  \bibinfo{author}{Dehghani, M.}, \bibinfo{author}{Minderer, M.},
  \bibinfo{author}{Heigold, G.}, \bibinfo{author}{Gelly, S.}, et~al.,
  \bibinfo{year}{2020}.
\newblock \bibinfo{title}{An image is worth 16x16 words: Transformers for image
  recognition at scale}.
\newblock \bibinfo{journal}{arXiv preprint arXiv:2010.11929} .
\bibitem[{Duan et~al.(2017)Duan, Liu, Jiao, Zhao and Zhang}]{duan2017sar}
\bibinfo{author}{Duan, Y.}, \bibinfo{author}{Liu, F.}, \bibinfo{author}{Jiao,
  L.}, \bibinfo{author}{Zhao, P.}, \bibinfo{author}{Zhang, L.},
  \bibinfo{year}{2017}.
\newblock \bibinfo{title}{Sar image segmentation based on convolutional-wavelet
  neural network and markov random field}.
\newblock \bibinfo{journal}{Pattern Recognition} \bibinfo{volume}{64},
  \bibinfo{pages}{255--267}.
\bibitem[{Everingham et~al.(2015)Everingham, Eslami, Van~Gool, Williams, Winn
  and Zisserman}]{Everingham15}
\bibinfo{author}{Everingham, M.}, \bibinfo{author}{Eslami, S.M.A.},
  \bibinfo{author}{Van~Gool, L.}, \bibinfo{author}{Williams, C.K.I.},
  \bibinfo{author}{Winn, J.}, \bibinfo{author}{Zisserman, A.},
  \bibinfo{year}{2015}.
\newblock \bibinfo{title}{The pascal visual object classes challenge: A
  retrospective}.
\newblock \bibinfo{journal}{International Journal of Computer Vision}
  \bibinfo{volume}{111}, \bibinfo{pages}{98--136}.
\bibitem[{Gong et~al.(2013)Gong, Zeng, Yan, Lv, Yang, Xu, Feng, Ding, Qi, Li
  et~al.}]{gong2013continuously}
\bibinfo{author}{Gong, H.}, \bibinfo{author}{Zeng, S.}, \bibinfo{author}{Yan,
  C.}, \bibinfo{author}{Lv, X.}, \bibinfo{author}{Yang, Z.},
  \bibinfo{author}{Xu, T.}, \bibinfo{author}{Feng, Z.}, \bibinfo{author}{Ding,
  W.}, \bibinfo{author}{Qi, X.}, \bibinfo{author}{Li, A.}, et~al.,
  \bibinfo{year}{2013}.
\newblock \bibinfo{title}{Continuously tracing brain-wide long-distance axonal
  projections in mice at a one-micron voxel resolution}.
\newblock \bibinfo{journal}{Neuroimage} \bibinfo{volume}{74},
  \bibinfo{pages}{87--98}.
\bibitem[{Guan et~al.(2019)Guan, Lai and Xiong}]{guan2019wavelet}
\bibinfo{author}{Guan, J.}, \bibinfo{author}{Lai, R.}, \bibinfo{author}{Xiong,
  A.}, \bibinfo{year}{2019}.
\newblock \bibinfo{title}{Wavelet deep neural network for stripe noise
  removal}.
\newblock \bibinfo{journal}{IEEE Access} \bibinfo{volume}{7},
  \bibinfo{pages}{44544--44554}.
\bibitem[{He et~al.(2016)He, Zhang, Ren and Sun}]{he2016deep}
\bibinfo{author}{He, K.}, \bibinfo{author}{Zhang, X.}, \bibinfo{author}{Ren,
  S.}, \bibinfo{author}{Sun, J.}, \bibinfo{year}{2016}.
\newblock \bibinfo{title}{Deep residual learning for image recognition}, in:
  \bibinfo{booktitle}{Proceedings of the IEEE conference on computer vision and
  pattern recognition}, pp. \bibinfo{pages}{770--778}.
\bibitem[{Huang et~al.(2017)Huang, He, Sun and Tan}]{huang2017wavelet}
\bibinfo{author}{Huang, H.}, \bibinfo{author}{He, R.}, \bibinfo{author}{Sun,
  Z.}, \bibinfo{author}{Tan, T.}, \bibinfo{year}{2017}.
\newblock \bibinfo{title}{Wavelet-srnet: A wavelet-based cnn for multi-scale
  face super resolution}, in: \bibinfo{booktitle}{Proceedings of the IEEE
  International Conference on Computer Vision}, pp.
  \bibinfo{pages}{1689--1697}.
\bibitem[{Huang et~al.(2020)Huang, Chen, Liu, Xu, Cao, Xu, Wang, Rao, Li, Zeng
  et~al.}]{huang2020weakly}
\bibinfo{author}{Huang, Q.}, \bibinfo{author}{Chen, Y.}, \bibinfo{author}{Liu,
  S.}, \bibinfo{author}{Xu, C.}, \bibinfo{author}{Cao, T.},
  \bibinfo{author}{Xu, Y.}, \bibinfo{author}{Wang, X.}, \bibinfo{author}{Rao,
  G.}, \bibinfo{author}{Li, A.}, \bibinfo{author}{Zeng, S.}, et~al.,
  \bibinfo{year}{2020}.
\newblock \bibinfo{title}{Weakly supervised learning of 3d deep network for
  neuron reconstruction}.
\newblock \bibinfo{journal}{Frontiers in Neuroanatomy} \bibinfo{volume}{14}.
\bibitem[{Jiang et~al.(2020)Jiang, Chen, Liu, Wang and Meijering}]{jiang20203d}
\bibinfo{author}{Jiang, Y.}, \bibinfo{author}{Chen, W.}, \bibinfo{author}{Liu,
  M.}, \bibinfo{author}{Wang, Y.}, \bibinfo{author}{Meijering, E.},
  \bibinfo{year}{2020}.
\newblock \bibinfo{title}{3d neuron microscopy image segmentation via the
  ray-shooting model and a dc-blstm network}.
\newblock \bibinfo{journal}{IEEE Transactions on Medical Imaging} .
\bibitem[{Kim and Pearlman(1997)}]{kim1997embedded}
\bibinfo{author}{Kim, B.J.}, \bibinfo{author}{Pearlman, W.A.},
  \bibinfo{year}{1997}.
\newblock \bibinfo{title}{An embedded wavelet video coder using
  three-dimensional set partitioning in hierarchical trees (spiht)}, in:
  \bibinfo{booktitle}{Proceedings DCC'97. Data Compression Conference},
  \bibinfo{organization}{IEEE}. pp. \bibinfo{pages}{251--260}.
\bibitem[{Klinghoffer et~al.(2020)Klinghoffer, Morales, Park, Evans, Chung and
  Brattain}]{klinghoffer2020self}
\bibinfo{author}{Klinghoffer, T.}, \bibinfo{author}{Morales, P.},
  \bibinfo{author}{Park, Y.G.}, \bibinfo{author}{Evans, N.},
  \bibinfo{author}{Chung, K.}, \bibinfo{author}{Brattain, L.J.},
  \bibinfo{year}{2020}.
\newblock \bibinfo{title}{Self-supervised feature extraction for 3d axon
  segmentation}, in: \bibinfo{booktitle}{Proceedings of the IEEE/CVF Conference
  on Computer Vision and Pattern Recognition Workshops}, pp.
  \bibinfo{pages}{978--979}.
\bibitem[{Ku et~al.(2016)Ku, Swaney, Park, Albanese, Murray, Cho, Park,
  Mangena, Chen and Chung}]{ku2016multiplexed}
\bibinfo{author}{Ku, T.}, \bibinfo{author}{Swaney, J.}, \bibinfo{author}{Park,
  J.Y.}, \bibinfo{author}{Albanese, A.}, \bibinfo{author}{Murray, E.},
  \bibinfo{author}{Cho, J.H.}, \bibinfo{author}{Park, Y.G.},
  \bibinfo{author}{Mangena, V.}, \bibinfo{author}{Chen, J.},
  \bibinfo{author}{Chung, K.}, \bibinfo{year}{2016}.
\newblock \bibinfo{title}{Multiplexed and scalable super-resolution imaging of
  three-dimensional protein localization in size-adjustable tissues}.
\newblock \bibinfo{journal}{Nature biotechnology} \bibinfo{volume}{34},
  \bibinfo{pages}{973--981}.
\bibitem[{Li et~al.(2010)Li, Gong, Zhang, Wang, Yan, Wu, Liu, Zeng and
  Luo}]{li2010micro}
\bibinfo{author}{Li, A.}, \bibinfo{author}{Gong, H.}, \bibinfo{author}{Zhang,
  B.}, \bibinfo{author}{Wang, Q.}, \bibinfo{author}{Yan, C.},
  \bibinfo{author}{Wu, J.}, \bibinfo{author}{Liu, Q.}, \bibinfo{author}{Zeng,
  S.}, \bibinfo{author}{Luo, Q.}, \bibinfo{year}{2010}.
\newblock \bibinfo{title}{Micro-optical sectioning tomography to obtain a
  high-resolution atlas of the mouse brain}.
\newblock \bibinfo{journal}{Science} \bibinfo{volume}{330},
  \bibinfo{pages}{1404--1408}.
\bibitem[{Li and Shen(2020)}]{li20193d}
\bibinfo{author}{Li, Q.}, \bibinfo{author}{Shen, L.}, \bibinfo{year}{2020}.
\newblock \bibinfo{title}{3d neuron reconstruction in tangled neuronal image
  with deep networks}.
\newblock \bibinfo{journal}{IEEE transactions on medical imaging}
  \bibinfo{volume}{39}, \bibinfo{pages}{425--435}.
\bibitem[{Li et~al.(2020)Li, Shen, Guo and Lai}]{qiufu_2020_CVPR}
\bibinfo{author}{Li, Q.}, \bibinfo{author}{Shen, L.}, \bibinfo{author}{Guo,
  S.}, \bibinfo{author}{Lai, Z.}, \bibinfo{year}{2020}.
\newblock \bibinfo{title}{Wavelet integrated cnns for noise-robust image
  classification}, in: \bibinfo{booktitle}{The IEEE Conference on Computer
  Vision and Pattern Recognition (CVPR)}.
\bibitem[{Li et~al.(2017)Li, Zeng, Peng and Ji}]{li2017deep}
\bibinfo{author}{Li, R.}, \bibinfo{author}{Zeng, T.}, \bibinfo{author}{Peng,
  H.}, \bibinfo{author}{Ji, S.}, \bibinfo{year}{2017}.
\newblock \bibinfo{title}{Deep learning segmentation of optical microscopy
  images improves 3-d neuron reconstruction}.
\newblock \bibinfo{journal}{IEEE transactions on medical imaging}
  \bibinfo{volume}{36}, \bibinfo{pages}{1533--1541}.
\bibitem[{Liu et~al.(2020)Liu, Liu, Yuan, Slabaugh, Leonardis, Zhou and
  Tian}]{liu2020wavelet}
\bibinfo{author}{Liu, L.}, \bibinfo{author}{Liu, J.}, \bibinfo{author}{Yuan,
  S.}, \bibinfo{author}{Slabaugh, G.}, \bibinfo{author}{Leonardis, A.},
  \bibinfo{author}{Zhou, W.}, \bibinfo{author}{Tian, Q.}, \bibinfo{year}{2020}.
\newblock \bibinfo{title}{Wavelet-based dual-branch network for image
  demoir{\'e}ing}.
\newblock \bibinfo{journal}{arXiv preprint arXiv:2007.07173} .
\bibitem[{Liu et~al.(2018a)Liu, Zhang, Zhang, Lin and Zuo}]{liu2018multi}
\bibinfo{author}{Liu, P.}, \bibinfo{author}{Zhang, H.}, \bibinfo{author}{Zhang,
  K.}, \bibinfo{author}{Lin, L.}, \bibinfo{author}{Zuo, W.},
  \bibinfo{year}{2018}a.
\newblock \bibinfo{title}{Multi-level wavelet-cnn for image restoration}, in:
  \bibinfo{booktitle}{Proceedings of the IEEE Conference on Computer Vision and
  Pattern Recognition Workshops}, pp. \bibinfo{pages}{773--782}.
\bibitem[{Liu et~al.(2016)Liu, Zhang, Liu, Feng, Peng and Cai}]{liu2016rivulet}
\bibinfo{author}{Liu, S.}, \bibinfo{author}{Zhang, D.}, \bibinfo{author}{Liu,
  S.}, \bibinfo{author}{Feng, D.}, \bibinfo{author}{Peng, H.},
  \bibinfo{author}{Cai, W.}, \bibinfo{year}{2016}.
\newblock \bibinfo{title}{Rivulet: 3d neuron morphology tracing with iterative
  back-tracking}.
\newblock \bibinfo{journal}{Neuroinformatics} \bibinfo{volume}{14},
  \bibinfo{pages}{387--401}.
\bibitem[{Liu et~al.(2018b)Liu, Zhang, Song, Peng and Cai}]{liu2018automated}
\bibinfo{author}{Liu, S.}, \bibinfo{author}{Zhang, D.}, \bibinfo{author}{Song,
  Y.}, \bibinfo{author}{Peng, H.}, \bibinfo{author}{Cai, W.},
  \bibinfo{year}{2018}b.
\newblock \bibinfo{title}{Automated 3-d neuron tracing with precise branch
  erasing and confidence controlled back tracking}.
\newblock \bibinfo{journal}{IEEE transactions on medical imaging}
  \bibinfo{volume}{37}, \bibinfo{pages}{2441--2452}.
\bibitem[{Magliaro et~al.(2017)Magliaro, Callara, Vanello and
  Ahluwalia}]{magliaro2017manual}
\bibinfo{author}{Magliaro, C.}, \bibinfo{author}{Callara, A.L.},
  \bibinfo{author}{Vanello, N.}, \bibinfo{author}{Ahluwalia, A.},
  \bibinfo{year}{2017}.
\newblock \bibinfo{title}{A manual segmentation tool for three-dimensional
  neuron datasets}.
\newblock \bibinfo{journal}{Frontiers in neuroinformatics}
  \bibinfo{volume}{11}, \bibinfo{pages}{36}.
\bibitem[{Narayanaswamy et~al.(2011)Narayanaswamy, Wang and
  Roysam}]{narayanaswamy20113}
\bibinfo{author}{Narayanaswamy, A.}, \bibinfo{author}{Wang, Y.},
  \bibinfo{author}{Roysam, B.}, \bibinfo{year}{2011}.
\newblock \bibinfo{title}{3-d image pre-processing algorithms for improved
  automated tracing of neuronal arbors}.
\newblock \bibinfo{journal}{Neuroinformatics} \bibinfo{volume}{9},
  \bibinfo{pages}{219--231}.
\bibitem[{Park et~al.(2019)Park, Sohn, Chen, McCue, Yun, Drummond, Ku, Evans,
  Oak, Trieu et~al.}]{park2019protection}
\bibinfo{author}{Park, Y.G.}, \bibinfo{author}{Sohn, C.H.},
  \bibinfo{author}{Chen, R.}, \bibinfo{author}{McCue, M.},
  \bibinfo{author}{Yun, D.H.}, \bibinfo{author}{Drummond, G.T.},
  \bibinfo{author}{Ku, T.}, \bibinfo{author}{Evans, N.B.},
  \bibinfo{author}{Oak, H.C.}, \bibinfo{author}{Trieu, W.}, et~al.,
  \bibinfo{year}{2019}.
\newblock \bibinfo{title}{Protection of tissue physicochemical properties using
  polyfunctional crosslinkers}.
\newblock \bibinfo{journal}{Nature biotechnology} \bibinfo{volume}{37},
  \bibinfo{pages}{73--83}.
\bibitem[{Paszke et~al.(2017)Paszke, Gross, Chintala, Chanan, Yang, DeVito,
  Lin, Desmaison, Antiga and Lerer}]{paszke2017automatic}
\bibinfo{author}{Paszke, A.}, \bibinfo{author}{Gross, S.},
  \bibinfo{author}{Chintala, S.}, \bibinfo{author}{Chanan, G.},
  \bibinfo{author}{Yang, E.}, \bibinfo{author}{DeVito, Z.},
  \bibinfo{author}{Lin, Z.}, \bibinfo{author}{Desmaison, A.},
  \bibinfo{author}{Antiga, L.}, \bibinfo{author}{Lerer, A.},
  \bibinfo{year}{2017}.
\newblock \bibinfo{title}{Automatic differentiation in pytorch} .
\bibitem[{Peng et~al.(2015)Peng, Hawrylycz, Roskams, Hill, Spruston, Meijering
  and Ascoli}]{peng2015bigneuron}
\bibinfo{author}{Peng, H.}, \bibinfo{author}{Hawrylycz, M.},
  \bibinfo{author}{Roskams, J.}, \bibinfo{author}{Hill, S.},
  \bibinfo{author}{Spruston, N.}, \bibinfo{author}{Meijering, E.},
  \bibinfo{author}{Ascoli, G.A.}, \bibinfo{year}{2015}.
\newblock \bibinfo{title}{Bigneuron: large-scale 3d neuron reconstruction from
  optical microscopy images}.
\newblock \bibinfo{journal}{Neuron} \bibinfo{volume}{87},
  \bibinfo{pages}{252--256}.
\bibitem[{Peng et~al.(2011)Peng, Long and Myers}]{peng2011automatic}
\bibinfo{author}{Peng, H.}, \bibinfo{author}{Long, F.}, \bibinfo{author}{Myers,
  G.}, \bibinfo{year}{2011}.
\newblock \bibinfo{title}{Automatic 3d neuron tracing using all-path pruning}.
\newblock \bibinfo{journal}{Bioinformatics} \bibinfo{volume}{27},
  \bibinfo{pages}{i239--i247}.
\bibitem[{Peng et~al.(2010a)Peng, Ruan, Atasoy and
  Sternson}]{peng2010automatic}
\bibinfo{author}{Peng, H.}, \bibinfo{author}{Ruan, Z.},
  \bibinfo{author}{Atasoy, D.}, \bibinfo{author}{Sternson, S.},
  \bibinfo{year}{2010}a.
\newblock \bibinfo{title}{Automatic reconstruction of 3d neuron structures
  using a graph-augmented deformable model}.
\newblock \bibinfo{journal}{Bioinformatics} \bibinfo{volume}{26},
  \bibinfo{pages}{i38--i46}.
\bibitem[{Peng et~al.(2010b)Peng, Ruan, Long, Simpson and Myers}]{peng2010v3d}
\bibinfo{author}{Peng, H.}, \bibinfo{author}{Ruan, Z.}, \bibinfo{author}{Long,
  F.}, \bibinfo{author}{Simpson, J.H.}, \bibinfo{author}{Myers, E.W.},
  \bibinfo{year}{2010}b.
\newblock \bibinfo{title}{V3d enables real-time 3d visualization and
  quantitative analysis of large-scale biological image data sets}.
\newblock \bibinfo{journal}{Nature biotechnology} \bibinfo{volume}{28},
  \bibinfo{pages}{348}.
\bibitem[{Quan et~al.(2016)Quan, Zhou, Li, Li, Li, Li, Lv, Luo, Gong and
  Zeng}]{quan2016neurogps}
\bibinfo{author}{Quan, T.}, \bibinfo{author}{Zhou, H.}, \bibinfo{author}{Li,
  J.}, \bibinfo{author}{Li, S.}, \bibinfo{author}{Li, A.}, \bibinfo{author}{Li,
  Y.}, \bibinfo{author}{Lv, X.}, \bibinfo{author}{Luo, Q.},
  \bibinfo{author}{Gong, H.}, \bibinfo{author}{Zeng, S.}, \bibinfo{year}{2016}.
\newblock \bibinfo{title}{Neurogps-tree: automatic reconstruction of
  large-scale neuronal populations with dense neurites}.
\newblock \bibinfo{journal}{Nature methods} \bibinfo{volume}{13},
  \bibinfo{pages}{51}.
\bibitem[{Ronneberger et~al.(2015)Ronneberger, Fischer and
  Brox}]{ronneberger2015u}
\bibinfo{author}{Ronneberger, O.}, \bibinfo{author}{Fischer, P.},
  \bibinfo{author}{Brox, T.}, \bibinfo{year}{2015}.
\newblock \bibinfo{title}{U-net: Convolutional networks for biomedical image
  segmentation}, in: \bibinfo{booktitle}{International Conference on Medical
  image computing and computer-assisted intervention},
  \bibinfo{organization}{Springer}. pp. \bibinfo{pages}{234--241}.
\bibitem[{Shi and Pun(2017)}]{shi20173d}
\bibinfo{author}{Shi, C.}, \bibinfo{author}{Pun, C.M.}, \bibinfo{year}{2017}.
\newblock \bibinfo{title}{3d multi-resolution wavelet convolutional neural
  networks for hyperspectral image classification}.
\newblock \bibinfo{journal}{Information Sciences} \bibinfo{volume}{420},
  \bibinfo{pages}{49--65}.
\bibitem[{Touvron et~al.(2020)Touvron, Cord, Douze, Massa, Sablayrolles and
  J{\'e}gou}]{touvron2020training}
\bibinfo{author}{Touvron, H.}, \bibinfo{author}{Cord, M.},
  \bibinfo{author}{Douze, M.}, \bibinfo{author}{Massa, F.},
  \bibinfo{author}{Sablayrolles, A.}, \bibinfo{author}{J{\'e}gou, H.},
  \bibinfo{year}{2020}.
\newblock \bibinfo{title}{Training data-efficient image transformers \&
  distillation through attention}.
\newblock \bibinfo{journal}{arXiv preprint arXiv:2012.12877} .
\bibitem[{Wang et~al.(2019a)Wang, Zhang, Song, Liu, Huang, Chen, Peng and
  Cai}]{wang2019multiscale}
\bibinfo{author}{Wang, H.}, \bibinfo{author}{Zhang, D.}, \bibinfo{author}{Song,
  Y.}, \bibinfo{author}{Liu, S.}, \bibinfo{author}{Huang, H.},
  \bibinfo{author}{Chen, M.}, \bibinfo{author}{Peng, H.}, \bibinfo{author}{Cai,
  W.}, \bibinfo{year}{2019}a.
\newblock \bibinfo{title}{Multiscale kernels for enhanced u-shaped network to
  improve 3d neuron tracing}, in: \bibinfo{booktitle}{Proceedings of the IEEE
  Conference on Computer Vision and Pattern Recognition Workshops}, pp.
  \bibinfo{pages}{0--0}.
\bibitem[{Wang et~al.(2019b)Wang, Li, Liu, Zhou, Ruan, Kong, Li, Wang, Zhong,
  Chai et~al.}]{wang2019teravr}
\bibinfo{author}{Wang, Y.}, \bibinfo{author}{Li, Q.}, \bibinfo{author}{Liu,
  L.}, \bibinfo{author}{Zhou, Z.}, \bibinfo{author}{Ruan, Z.},
  \bibinfo{author}{Kong, L.}, \bibinfo{author}{Li, Y.}, \bibinfo{author}{Wang,
  Y.}, \bibinfo{author}{Zhong, N.}, \bibinfo{author}{Chai, R.}, et~al.,
  \bibinfo{year}{2019}b.
\newblock \bibinfo{title}{Teravr empowers precise reconstruction of complete
  3-d neuronal morphology in the whole brain}.
\newblock \bibinfo{journal}{Nature communications} \bibinfo{volume}{10},
  \bibinfo{pages}{1--9}.
\bibitem[{Wang et~al.(2011)Wang, Narayanaswamy, Tsai and
  Roysam}]{wang2011broadly}
\bibinfo{author}{Wang, Y.}, \bibinfo{author}{Narayanaswamy, A.},
  \bibinfo{author}{Tsai, C.L.}, \bibinfo{author}{Roysam, B.},
  \bibinfo{year}{2011}.
\newblock \bibinfo{title}{A broadly applicable 3-d neuron tracing method based
  on open-curve snake}.
\newblock \bibinfo{journal}{Neuroinformatics} \bibinfo{volume}{9},
  \bibinfo{pages}{193--217}.
\bibitem[{Williams and Li(2018)}]{williams2018wavelet}
\bibinfo{author}{Williams, T.}, \bibinfo{author}{Li, R.}, \bibinfo{year}{2018}.
\newblock \bibinfo{title}{Wavelet pooling for convolutional neural networks},
  in: \bibinfo{booktitle}{International Conference on Learning
  Representations}.
\bibitem[{Wu et~al.(2009)Wu, Jiang, Fang, Jiao and Shi}]{wu2009hyperspectral}
\bibinfo{author}{Wu, J.}, \bibinfo{author}{Jiang, K.}, \bibinfo{author}{Fang,
  Y.}, \bibinfo{author}{Jiao, L.}, \bibinfo{author}{Shi, G.},
  \bibinfo{year}{2009}.
\newblock \bibinfo{title}{Hyperspectral image compression using distributed
  source coding and 3d speck}, in: \bibinfo{booktitle}{MIPPR 2009:
  Multispectral Image Acquisition and Processing},
  \bibinfo{organization}{International Society for Optics and Photonics}. p.
  \bibinfo{pages}{74940Z}.
\bibitem[{Xiao and Peng(2013)}]{xiao2013app2}
\bibinfo{author}{Xiao, H.}, \bibinfo{author}{Peng, H.}, \bibinfo{year}{2013}.
\newblock \bibinfo{title}{App2: automatic tracing of 3d neuron morphology based
  on hierarchical pruning of a gray-weighted image distance-tree}.
\newblock \bibinfo{journal}{Bioinformatics} \bibinfo{volume}{29},
  \bibinfo{pages}{1448--1454}.
\bibitem[{Yang et~al.(2019)Yang, Zhao, Chan and Xiao}]{yang2019multi}
\bibinfo{author}{Yang, J.}, \bibinfo{author}{Zhao, Y.Q.},
  \bibinfo{author}{Chan, J.C.W.}, \bibinfo{author}{Xiao, L.},
  \bibinfo{year}{2019}.
\newblock \bibinfo{title}{A multi-scale wavelet 3d-cnn for hyperspectral image
  super-resolution}.
\newblock \bibinfo{journal}{Remote sensing} \bibinfo{volume}{11},
  \bibinfo{pages}{1557}.
\bibitem[{Yoo et~al.(2019)Yoo, Uh, Chun, Kang and Ha}]{yoo2019photorealistic}
\bibinfo{author}{Yoo, J.}, \bibinfo{author}{Uh, Y.}, \bibinfo{author}{Chun,
  S.}, \bibinfo{author}{Kang, B.}, \bibinfo{author}{Ha, J.W.},
  \bibinfo{year}{2019}.
\newblock \bibinfo{title}{Photorealistic style transfer via wavelet
  transforms}, in: \bibinfo{booktitle}{Proceedings of the IEEE International
  Conference on Computer Vision}, pp. \bibinfo{pages}{9036--9045}.
\bibitem[{Zhang(2019)}]{zhang2019making}
\bibinfo{author}{Zhang, R.}, \bibinfo{year}{2019}.
\newblock \bibinfo{title}{Making convolutional networks shift-invariant again},
  in: \bibinfo{booktitle}{Proceedings of the International Conference on
  Machine Learning}, pp. \bibinfo{pages}{7324--7334}.
\bibitem[{Zhou et~al.(2016)Zhou, Liu, Long and Peng}]{zhou2016tremap}
\bibinfo{author}{Zhou, Z.}, \bibinfo{author}{Liu, X.}, \bibinfo{author}{Long,
  B.}, \bibinfo{author}{Peng, H.}, \bibinfo{year}{2016}.
\newblock \bibinfo{title}{Tremap: automatic 3d neuron reconstruction based on
  tracing, reverse mapping and assembling of 2d projections}.
\newblock \bibinfo{journal}{Neuroinformatics} \bibinfo{volume}{14},
  \bibinfo{pages}{41--50}.
\bibitem[{Zou et~al.(2020)Zou, Xiao, Yu and Lee}]{zou2020delving}
\bibinfo{author}{Zou, X.}, \bibinfo{author}{Xiao, F.}, \bibinfo{author}{Yu,
  Z.}, \bibinfo{author}{Lee, Y.J.}, \bibinfo{year}{2020}.
\newblock \bibinfo{title}{Delving deeper into anti-aliasing in convnets}, in:
  \bibinfo{booktitle}{Proceedings of the British Machine Vision Conference}.

\end{thebibliography}

\clearpage\begin{strip}
\begin{align}
\nonumber
\large
\mbox{\textbf{Supplementary for ``Neuron Segmentation using 3D Wavelet Integrated Encoder-Decoder Network''}}
\end{align}
\end{strip}

\renewcommand\thesection{S.\arabic{section}}
\renewcommand\theequation{S.\arabic{equation}}
\renewcommand\thetable{S.\arabic{table}}
\renewcommand\thefigure{S.\arabic{figure}}
\setcounter{section}{0}
\setcounter{equation}{0}
\setcounter{table}{0}
\setcounter{figure}{0}
\section{3D wavelet filters}
\label{sec_3D_wavelet_filters}
Generally, the 3D wavelet filters are tensor products of the two filters of 1D wavelet, $\text{f}_l, \text{f}_h$, i.e.,
\begin{align}
\label{eq_3D_filter_from_1D_filter}
\text{f}_{c_0c_1c_2} &= \text{f}_{c_0} \otimes\text{f}_{c_1} \otimes \text{f}_{c_2},
\quad c_0, c_1, c_2  \in \{l, h\},
\end{align}
where $\otimes$ represents the tensor product.
Take Haar wavelet for example,
the low-pass and high-pass filters of 1D Haar wavelet are
\begin{align}
	\text{f}^{\text{H}}_l = \frac{1}{\sqrt{2}}(1,1)^T,\quad \text{f}^{\text{H}}_h = \frac{1}{\sqrt{2}}(1,-1)^T.
\end{align}
Then, via Eq. (\ref{eq_3D_filter_from_1D_filter}), we get the filters of the corresponding 3D Haar wavelet:
\begin{align}
\text{f}_{lll}^{\text{H}} &
=\frac{1}{2\sqrt{2}}\left[
\left[\begin{array}{cc}
1&1\\
1&1
\end{array}\right];
\left[\begin{array}{cc}
1&1\\
1&1
\end{array}\right]
\right],\\
\text{f}_{llh}^{\text{H}} &
=\frac{1}{2\sqrt{2}}\left[
\left[\begin{array}{cc}
1&-1\\
1&-1
\end{array}\right];
\left[\begin{array}{cc}
1&-1\\
1&-1
\end{array}\right]
\right],\\
\text{f}_{lhl}^{\text{H}} &
=\frac{1}{2\sqrt{2}}\left[
\left[\begin{array}{cc}
1&1\\
-1&-1
\end{array}\right];
\left[\begin{array}{cc}
1&1\\
-1&-1
\end{array}\right]
\right],\\
\text{f}_{lhh}^{\text{H}} &
=\frac{1}{2\sqrt{2}}\left[
\left[\begin{array}{cc}
1&-1\\
-1&1
\end{array}\right];
\left[\begin{array}{cc}
1&-1\\
-1&1
\end{array}\right]
\right],\\
\text{f}_{hll}^{\text{H}} &
=\frac{1}{2\sqrt{2}}\left[
\left[\begin{array}{cc}
1&1\\
1&1
\end{array}\right];
\left[\begin{array}{cc}
-1&-1\\
-1&-1
\end{array}\right]
\right],\\
\text{f}_{hlh}^{\text{H}} &
=\frac{1}{2\sqrt{2}}\left[
\left[\begin{array}{cc}
1&-1\\
1&-1
\end{array}\right];
\left[\begin{array}{cc}
-1&1\\
-1&1
\end{array}\right]
\right],\\
\text{f}_{hhl}^{\text{H}} &
=\frac{1}{2\sqrt{2}}\left[
\left[\begin{array}{cc}
1&1\\
-1&-1
\end{array}\right];
\left[\begin{array}{cc}
-1&-1\\
1&1
\end{array}\right]
\right],\\
\text{f}_{hhh}^{\text{H}} &
=\frac{1}{2\sqrt{2}}\left[
\left[\begin{array}{cc}
1&-1\\
-1&1
\end{array}\right];
\left[\begin{array}{cc}
-1&1\\
1&-1
\end{array}\right]
\right].
\end{align}

We here introduce the commonly used orthogonal Daubechies wavelets and biorthogonal Cohen wavelets.
Their corresponding 3D wavelet filters are designed according to Eq. (\ref{eq_3D_filter_from_1D_filter}).

\textbf{Orthogonal wavelets}\label{APP_ortho_wavelet}\quad
Daubechies wavelet is orthogonal, a set of orthogonal basis for $L^2(x)$ could be derived from its scaling and wavelet functions.
Daubechies wavelet has an approximation order parameter $p$,
and the length of its filter is $2p$.
Table \ref{Tab_Daubechies_banks} shows the low-pass filter $\text{f}_l = \{f_k^{(l)}\}$ of the wavelets with order $p, 1\leq p\leq6$,
while the high-pass filter $\text{f}_h = \{f_k^{(h)}\}$ can be deduced from
\begin{equation}\label{eq_high_pass_bank}
f_k^{(h)} = (-1)^k f^{(l)}_{\mathcal{N}-k},
\end{equation}
where $\mathcal{N}$ is an odd number.
Daubechies(1) is Haar wavelet.

\textbf{Biorthogonal wavelets}\label{APP_bior_wavelet}\quad
Cohen wavelets are symmetric biorthogonal wavelets,
and each of them is associated with scaling function $\phi$, wavelet function $\psi$, and their dual functions $\tilde{\phi}, \tilde{\psi}$.
Correspondingly, it has four filters $\text{f}_l$, $\text{f}_h$, $\tilde{\text{f}}_l$, and $\tilde{\text{f}}_h$.
While a signal is decomposed using filters $\text{f}_l$ and $\text{f}_h$ with DWT,
it can be reconstructed using the dual filters $\tilde{\text{f}}_l$ and $\tilde{\text{f}}_h$ with IDWT.
Cohen wavelet is with two order parameters $p$ and $\tilde{p}$.
Table \ref{Tab_CDF_banks} shows the low-pass filters with orders $2\leq p = \tilde{p}\leq5$.
Their high-pass filters can be deduced from
\begin{eqnarray}
\label{eq_high_pass_bank_bior_1}
&f_k^{(h)} = (-1)^k \tilde{f}_{\mathcal{N}-k}^{(l)},&\\
\label{eq_high_pass_bank_bior_2}
&\tilde{f}_k^{(h)} = (-1)^k f_{\mathcal{N}-k}^{(l)},&
\end{eqnarray}
where $\mathcal{N}$ is an odd number.
Cohen$(1,1)$ is Haar wavelet.

Wavelet theory is valid for finite or infinite filters,
but the infinite case is rarely covered in practical interest.

\begin{table*}
	\caption{Low-pass filters of the Daubechies wavelets.
    {\color{black}The high-pass filters of Daubechies wavelets could be deduced from the low-pass filters via Eq. (\ref{eq_high_pass_bank}).
    Daubechies wavelets are orthogonal.}}
	\label{Tab_Daubechies_banks}
	\begin{center}
	\setlength{\tabcolsep}{1mm}{
	\begin{tabular}{c|cccccc}
		\hline
		          $p$           &     $1$      &       $2$       &         $3$         &         $4$         &         $5$         &         $6$         \\ \hline
		\multirow{12}{*}{$f_k^{(l)}$} &     $1$      &  $1+\sqrt{3}$   & $~~~0.332670552950$ & $~~~0.230377813309$ & $~~~0.160102397974$ & $~~~0.111540743350$ \\
		                        &     $1$      &  $3+\sqrt{3}$   & $~~~0.806891509311$ & $~~~0.714846570553$ & $~~~0.603829269797$ & $~~~0.494623890398$ \\
		                        &              &  $3-\sqrt{3}$   & $~~~0.459877502118$ & $~~~0.630880767930$ & $~~~0.724308528438$ & $~~~0.751133908021$ \\
		                        &              &  $1-\sqrt{3}$   &  $-0.135011020010$  &  $-0.027983769417$  & $~~~0.138428145901$ & $~~~0.315250351709$ \\
		                        &              &                 &  $-0.085441273882$  &  $-0.187034811719$  &  $-0.242294887066$  &  $-0.226264693965$  \\
		                        &              &                 & $~~~0.035226291886$ & $~~~0.030841381836$ &  $-0.032244869585$  &  $-0.129766867567$  \\
		                        &              &                 &                     & $~~~0.032883011667$ & $~~~0.077571493840$ & $~~~0.097501605587$ \\
		                        &              &                 &                     &  $-0.010597401785$  &  $-0.006241490213$  & $~~~0.027522865530$ \\
		                        &              &                 &                     &                     &  $-0.012580751999$  &  $-0.031582039317$  \\
		                        &              &                 &                     &                     & $~~~0.003335725285$ & $~~~0.000553842201$ \\
		                        &              &                 &                     &                     &                     & $~~~0.004777257511$ \\
		                        &              &                 &                     &                     &                     &  $-0.001077301085$  \\ \hline
		       $\text{factor}$         & $1/\sqrt{2}$ & $1/(4\sqrt{2})$ &         $1$         &         $1$         &         $1$         &         $1$         \\ \hline
	\end{tabular}}
	\end{center}
\end{table*}
\begin{table*}
\caption{Low-pass filters of the Cohen wavelets.
{\color{black}The high-pass filters of Cohen wavelets could be deduced via Eqs. (\ref{eq_high_pass_bank_bior_1})-(\ref{eq_high_pass_bank_bior_2}).
The filters and dual filters of biorthogonal Cohen wavelets are applied to decompose and reconstruct image, respectively.}}
\label{Tab_CDF_banks}
\begin{center}
	\setlength{\tabcolsep}{0.5mm}{
		\begin{tabular}{c|cc|cc|cc|cc}
			\hline
			$(p,\tilde{p})$     & \multicolumn{2}{|c|}{$(2,2)$}       & \multicolumn{2}{|c|}{$(3,3)$}       & \multicolumn{2}{|c|}{$(4,4)$}          & \multicolumn{2}{|c}{$(5,5)$}           \\ \hline
			filter          & $\text{f}_l$ & $\tilde{\text{f}}_l$ & $\text{f}_l$ & $\tilde{\text{f}}_l$ &  $\text{f}_l$   & $\tilde{\text{f}}_l$ &  $\text{f}_l$   & $\tilde{\text{f}}_l$ \\ \hline
			\multirow{12}{*}{$f_k^{(l)}$} &     $0$      &         $0$          &     $0$      &   $~~~0.06629126$    &       $0$       &         $0$          & $~~~0.01345671$ &         $0$          \\
			& $0.35355339$ &    $-0.17677670$     &     $0$      &    $-0.19887378$     &  $-0.06453888$  &   $~~~0.03782846$    &  $-0.00269497$  &         $0$          \\
			& $0.70710678$ &   $~~~0.35355339$    & $0.17677670$ &    $-0.15467961$     &  $-0.04068942$  &    $-0.02384947$     &  $-0.13670658$  &   $~~~0.03968709$    \\
			& $0.35355339$ &   $~~~1.06066017$    & $0.53033009$ &   $~~~0.99436891$    & $~~~0.41809227$ &    $-0.11062440$     &  $-0.09350470$  &   $~~~0.00794811$    \\
			&     $0$      &   $~~~0.35355339$    & $0.53033009$ &   $~~~0.99436891$    & $~~~0.78848562$ &   $~~~0.37740286$    & $~~~0.47680327$ &    $-0.05446379$     \\
			&     $0$      &    $-0.17677670$     & $0.17677670$ &    $-0.15467961$     & $~~~0.41809227$ &   $~~~0.85269868$    & $~~~0.89950611$ &   $~~~0.34560528$    \\
			&              &                      &     $0$      &    $-0.19887378$     &  $-0.04068942$  &   $~~~0.37740286$    & $~~~0.47680327$ &   $~~~0.73666018$    \\
			&              &                      &     $0$      &   $~~~0.06629126$    &  $-0.06453888$  &    $-0.11062440$     &  $-0.09350470$  &   $~~~0.34560528$    \\
			&              &                      &              &                      &       $0$       &    $-0.02384947$     &  $-0.13670658$  &    $-0.05446379$     \\
			&              &                      &              &                      &       $0$       &   $~~~0.03782846$    &  $-0.00269497$  &   $~~~0.00794811$    \\
			&              &                      &              &                      &                 &                      & $~~~0.01345671$ &   $~~~0.03968709$    \\
			&              &                      &              &                      &                 &                      &       $0$       &         $0$          \\ \hline
	\end{tabular}}
\end{center}
\end{table*}

\section{The naive down-sampling and up-sampling}
\label{sec_naive_sampling}
In Eqs. (\ref{eq_DWT}) and (\ref{eq_IDWT}), 3D DWT and IDWT are implemented using 3D naive down-sampling and up-sampling.
For a tensor $\textbf{\emph{X}} =\{X_{i,j,k}\}\in\mathbb{R}^{d\times m\times n}$,
\begin{equation}\label{eq_naive_down_sampling}
(\downarrow2)\textbf{\emph{X}} = \{((\downarrow2)\textbf{\emph{X}})_{i,j,k}\} \in \mathbb{R}^{\lfloor\frac{d}{2}\rfloor\times\lfloor\frac{m}{2}\rfloor\times\lfloor\frac{n}{2}\rfloor},
\end{equation}
and
\begin{equation}\label{eq_naive_down_sampling_index}
 ((\downarrow2)\textbf{\emph{X}})_{i,j,k} = X_{2i,2j,2k}.
\end{equation}

For a 3D tensor
$\textbf{\emph{X}} =\{X_{i,j,k}\}\in\mathbb{R}^{d\times m\times n}$,
\begin{equation}\label{eq_naive_up_sampling}
(\uparrow2)\textbf{\emph{X}} = \{((\uparrow2)\textbf{\emph{X}})_{i,j,k}\}\in\mathbb{R}^{2d\times 2m\times 2n},
\end{equation}
and
\begin{equation}\label{eq_naive_up_sampling_index}
((\uparrow2)\textbf{\emph{X}})_{i,j,k} = \left\{\begin{array}{cc}X_{\frac{i}{2},\frac{j}{2},\frac{k}{2}}&\text{if}~\frac{i}{2},\frac{j}{2},\frac{k}{2} \in\mathbb{Z},\\0&\text{else}.\end{array}\right.
\end{equation}

\section{The configuration for 3D WaveUNet}
\label{sec_configuration_3D_WaveUNet}
\begin{table}[!t]
	\scriptsize
    \begin{center}
	\caption{Deep network configurations.}
	\label{tab_network_configuration}
		\setlength{\tabcolsep}{1.75mm}
			\begin{tabular}{c||l|r||c|c}\cline{1-5}
				\multirow{2}{*}{data size} &    \multicolumn{2}{c||}{channel number} & \multirow{2}{*}{3D U-Nets$^a$} & \multirow{2}{*}{3D WaveUNets$^b$}\\ \cline{2-3}
				                           & \multicolumn{1}{c|}{encoder} & \multicolumn{1}{c||}{decoder}&  &\\\cline{1-5}
				      $32\times128\times128$& 1, 4              &4, 4   &DS-$x$&WADS-$y$\\
				      $16\times64\times64$  & 4, 8              &8, 4   &DS-$x$&WADS-$y$\\
				      $8\times32\times32$   & 8, 16            &16, 8 &DS-$x$&WADS-$y$\\
				      $4\times16\times16$   & 16, 32            &32, 16 &DS-$x$&WADS-$y$\\\cdashline{1-5}[2.25pt/3pt]
				      $4\times16\times16$   &  \multicolumn{2}{c||}{32, 32}&  \multicolumn{2}{c}{bottom block} \\\cline{1-5}
			\end{tabular}

    \end{center}
    \par
    \hspace{0pt}$^a$ The three 3D U-Nets are named as 3D U-Net($x$), $x \in \{\text{PU, PDc, ScIn}\}$.\par
    \hspace{0pt}$^b$ The four 3D WaveUNets designed in this paper are named as 3D WaveUNet($y$), $y \in \{\text{DDc, DIn, DI, DIDn}\}$.
\end{table}

In Sec. \ref{sec_3D_waveunet}, using the seven dual structures, we design seven 3D encoder-decoder networks for neuron segmentation.
While the first three networks are variants of 3D U-Net, the last four are 3D WaveUNets designed in this paper.
Each of them contains four nested dual structures with one bottom block containing two convolutions.
Table \ref{tab_network_configuration} illustrates their configurations.
In Table \ref{tab_network_configuration}, the first column shows the input size.
Every number in the table corresponds to a convolutional layer with batch normalization (BN) and ReLU.
While the number in column ``encoder'' is the number of input channels of the convolution,
the number in column ``decoder'' is that of output channels.
At the end of networks, a convolution with kernel size of $1\times1$ converts the output of decoder into the predicted segmentation result.

\subsection{The denoising block used in 3D WaveUNet(DIDn)}
The denoising block, used in 3D WaveUNet(DIDn), is implemented by hard shrinkage with threshold $\lambda = 0.25$:
\begin{align}
\label{eq_hard_threshold}
\text{HardShrink}(x) = \begin{cases}
        x, & \text{ if } x > \lambda , \\
        x, & \text{ if } x < -\lambda, \\
        0, & \text{otherwise}.
        \end{cases}
\end{align}
In the denoising block, we filter every coefficient $x$ in the seven high-frequency components
$$\textbf{\emph{X}}_c,~c\in\{llh, lhl, lhh, hll, hlh, hhl, hhh\},$$
according to Eq. (\ref{eq_hard_threshold}).

\section{3D neuron reconstruction using various tracing algorithms}
\label{sec_results_various_tracing_algorithms}
{\color{black}
To better illustrate the effectiveness of 3D wavelet integrated deep networks on 3D neuron reconstruction,
we apply various automatic tracing approaches to reconstruct the test images segmented by the different 3D deep networks integrated with or without 3D wavelets.
Table \ref{Tab_segmentation_reconstruction_5} shows the reconstruction performances of five reconstruction approaches,
including APP2 \citep{xiao2013app2}, Mean-shift Spanning Tree (MST) tracer \citep{peng2010v3d}, NeuroGPS \citep{quan2016neurogps},
Snake tracer \citep{narayanaswamy20113}, and TReMAP \citep{zhou2016tremap}, on the 28 test neuronal images segmented by various 3D deep networks.
From Table \ref{Tab_segmentation_reconstruction_5}, one can find that,
although these automatic tracing approaches perform diversely on the segmented neuronal images,
3D wavelet could consistently improve the reconstruction performance of these tracing approaches.
}
\begin{table*}
\scriptsize
    \caption{Reconstruction results of various automatic tracing approaches on neuronal images segmented by 3D deep networks integrated with or without 3D wavelets.}\label{Tab_segmentation_reconstruction_5}
    \begin{center}
    \setlength{\tabcolsep}{1.2mm}{
    \begin{tabular}{c|c||ccc|ccc|ccc|ccc|ccc}\cline{1-17}
    \multicolumn{2}{c||}{3D Network}&\multicolumn{15}{c}{Reconstruction}\\\cline{1-17}
    \multirow{2}{*}{Architecture } &Dual& \multicolumn{3}{c|}{APP2}& \multicolumn{3}{c|}{MST}& \multicolumn{3}{c|}{NeuroGPSTree}& \multicolumn{3}{c|}{Snake tracer}& \multicolumn{3}{c}{TReMap} \\\cline{3-17}
                            & Structure  &ESA&DSA&PDS&ESA&DSA&PDS&ESA&DSA&PDS&ESA&DSA&PDS&ESA&DSA&PDS  \\\cline{1-17}
    \multirow{3}{*}{3D U-Net}&DS-PU     &2.5444 &    7.3450 &    0.2200  &  5.2071  & 13.6657  &   0.2840&3.3983 &    7.4945  &   0.3475
                & 5.4563 &   13.5233 &    0.2703&4.9636 &   13.5156  &   0.2909\\
                             &DS-PDc    &2.4046 &    7.1799 &    0.1950  &4.9122    & 13.8708  &   0.2674&3.3042 &    7.5735  &   0.3300
                & 5.3888 &   13.6766 &    0.2644&4.6019 &   13.5037  &   0.2627\\
                             &DS-ScIn   &2.3614 &    6.9439 &    0.1968  &4.8481    & 13.5585  &   0.2650&3.3046 &    7.7271  &   0.3328
                & 6.4323 &   14.8387 &    0.2620&4.7483 &   13.5524  &   0.2688\\\cline{1-17}
    \multirow{4}{*}{3D WaveUNet}
                             &WADS-DDc  &2.2495 &    6.5712 &    0.1956  & \textbf{4.5830}   &  \textbf{12.9024}  &   \textbf{0.2578}&3.1746 &    7.3695 &    \textbf{0.3287}
                & 5.3404 &   13.5490 &    0.2645&4.3416 &   12.7909  &   0.2650\\
                             &WADS-DIn  &2.0288 &    6.2569 &    0.1922  &4.8316    &  13.3179  &   0.2668&3.2134 &    7.4323 &    0.3350
                & \textbf{5.2720} &   \textbf{13.3878} &    0.2644&4.2825 &   12.4185  &   0.2658\\
                             &WADS-DI   &2.0676 &   6.2566  &   \textbf{0.1857}     &  4.8255   &   13.6180 &  0.2612&\textbf{3.1214}    & \textbf{7.3349} &    0.3358
                & 5.3402 &  13.5475  &   \textbf{0.2605}&\textbf{4.0762}  &  \textbf{12.1576}  &   \textbf{0.2599}\\
                             &WADS-DIDn &\textbf{1.9973}    &  \textbf{6.0173}      &  0.1897   &   4.8720  &  13.5247  &   0.2675&3.2096  &   7.4646  &   0.3384
                & 5.4028 &   13.5505 &    0.2638&4.2252 &   12.5162  &   0.2624\\\cline{1-17}
    \end{tabular}}
    \end{center}
\end{table*}

\begin{figure}[tbp]
\color{black}
	\centering
	\subfigure[neuronal cube.]
	{\includegraphics*[scale=0.775, viewport=141 446 244 680]{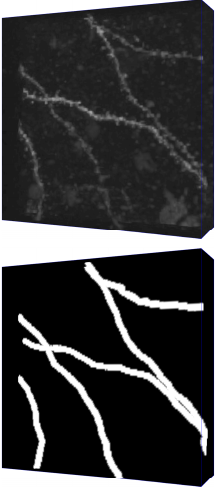}
	\label{fig_neuronal_natural_a}}\hspace{10pt}
	\subfigure[natural image.]
	{\includegraphics*[scale=0.775, viewport=282 446 455 680]{figures/natural_image_neuronal_cube.pdf}
	\label{fig_neuronal_natural_b}}
    \caption{Comparison of neuronal image and natural image.
            (a) A neuronal cube (top) with its label matrix (bottom).
            (b) An example image (top) with its manual segmentation result (bottom).}
	\label{fig_neuronal_natural}
\end{figure}
\section{\color{black}Comparison of neuronal image and natural image}
{\color{black}
Different from the common objects in natural images, the nerve fibers in neuronal cubes are line-shaped, with zigzag edges.
Fig. \ref{fig_neuronal_natural} shows an example neural cube from NeuCuDa and an example image from Pascal VOC \citep{Everingham15}.
During the annotation of nerve fiber in the cube, some label noises are introduced into the label matrix, because of the zigzag fiber edges.
The label noises occupy a significant proportion in the voxels labeled as ``nerve fiber'';
in the natural images, although some label noises also appear near the object edges, they only occupy a small proportion.
Therefore, the experiences sourced from natural image might not apply to the neuronal images, due to the gap between two domains.
}

\end{document}